\documentclass[letterpaper, twocolumn]{IEEEtran}

\usepackage[mathscr]{eucal}
\usepackage[cmex10]{amsmath}
\usepackage{epsfig,epsf,psfrag}
\usepackage{amssymb,amsmath,amsthm,amsfonts,latexsym}
\usepackage{amsmath,graphicx,bm,xcolor,url}
\usepackage{fixltx2e}
\usepackage{array}
\usepackage{verbatim}
\usepackage{bm}
\usepackage{algorithmic, cite}
\usepackage{algorithm}
\usepackage{verbatim}
\usepackage{textcomp}
\usepackage{mathrsfs}
\usepackage{epstopdf}

\catcode`~=11 \def\UrlSpecials{\do\~{\kern -.15em\lower .7ex\hbox{~}\kern .04em}} \catcode`~=13 

\allowdisplaybreaks[4]
 
\newcommand{\nn}{\nonumber}

\newcommand{\calD}{\mathcal{D}}

\newcommand{\calK}{\mathcal{K}}
\newcommand{\calL}{\mathcal{L}}
\newcommand{\calM}{\mathcal{M}}

\newcommand{\calX}{\mathcal{X}}
\newcommand{\calY}{\mathcal{Y}}
\newcommand{\calZ}{\mathcal{Z}}

\newcommand{\bp}{\mathbf{p}}

\newcommand{\rmb}{\mathrm{b}}

\newcommand{\rmd}{\mathrm{d}}

\newcommand{\rme}{\mathrm{e}}

\newcommand{\rmQ}{\mathrm{Q}}

\newcommand{\bbE}{\mathbb{E}}
\newcommand{\bbF}{\mathbb{F}}

\newcommand{\bbR}{\mathbb{R}}

\newcommand{\bbV}{\mathbb{V}}

\newcommand{\frakC}{\mathfrak{C}}

\DeclareMathAlphabet{\mathbsf}{OT1}{cmss}{bx}{n}
\DeclareMathAlphabet{\mathssf}{OT1}{cmss}{m}{sl}

\newcommand{\rvH}{\mathsf{H}}

\DeclareSymbolFont{bsfletters}{OT1}{cmss}{bx}{n}  
\DeclareSymbolFont{ssfletters}{OT1}{cmss}{m}{n}
\DeclareMathSymbol{\bsfGamma}{0}{bsfletters}{'000}
\DeclareMathSymbol{\ssfGamma}{0}{ssfletters}{'000}
\DeclareMathSymbol{\bsfDelta}{0}{bsfletters}{'001}
\DeclareMathSymbol{\ssfDelta}{0}{ssfletters}{'001}
\DeclareMathSymbol{\bsfTheta}{0}{bsfletters}{'002}
\DeclareMathSymbol{\ssfTheta}{0}{ssfletters}{'002}
\DeclareMathSymbol{\bsfLambda}{0}{bsfletters}{'003}
\DeclareMathSymbol{\ssfLambda}{0}{ssfletters}{'003}
\DeclareMathSymbol{\bsfXi}{0}{bsfletters}{'004}
\DeclareMathSymbol{\ssfXi}{0}{ssfletters}{'004}
\DeclareMathSymbol{\bsfPi}{0}{bsfletters}{'005}
\DeclareMathSymbol{\ssfPi}{0}{ssfletters}{'005}
\DeclareMathSymbol{\bsfSigma}{0}{bsfletters}{'006}
\DeclareMathSymbol{\ssfSigma}{0}{ssfletters}{'006}
\DeclareMathSymbol{\bsfUpsilon}{0}{bsfletters}{'007}
\DeclareMathSymbol{\ssfUpsilon}{0}{ssfletters}{'007}
\DeclareMathSymbol{\bsfPhi}{0}{bsfletters}{'010}
\DeclareMathSymbol{\ssfPhi}{0}{ssfletters}{'010}
\DeclareMathSymbol{\bsfPsi}{0}{bsfletters}{'011}
\DeclareMathSymbol{\ssfPsi}{0}{ssfletters}{'011}
\DeclareMathSymbol{\bsfOmega}{0}{bsfletters}{'012}
\DeclareMathSymbol{\ssfOmega}{0}{ssfletters}{'012}

\newcommand{\hatP}{\hat{P}}
\newcommand{\tilp}{\tilde{p}}
\newcommand{\tilP}{\tilde{P}}

\newcommand{\hatQ}{\hat{Q}}

\newcommand{\tilQ}{\tilde{Q}}

\newcommand{\hatW}{\hat{W}}

\newcommand{\barP}{\bar{P}}
\newcommand{\barQ}{\bar{Q}}

\DeclareMathOperator{\supp}{supp}

\newcommand{\eps}{\varepsilon}

\newcommand{\markov}{\mathrel{\multimap}\joinrel\mathrel{-}%
\joinrel\mathrel{\mkern-6mu}\joinrel\mathrel{-}}

\newtheorem{theorem}{Theorem} 
\newtheorem{condition}{Condition} 

\newtheorem{lemma}{Lemma}

\newtheorem{corollary}{Corollary}
\newtheorem{definition}{Definition}

\newcommand{\qednew}{\nobreak \ifvmode \relax \else
      \ifdim\lastskip<1.5em \hskip-\lastskip
      \hskip1.5em plus0em minus0.5em \fi \nobreak
      \vrule height0.75em width0.5em depth0.25em\fi}

\usepackage{cite, enumitem}

\begin{document}
\flushbottom
\allowdisplaybreaks[1]

\title{Time-Division is Optimal for Covert Communication over Some     Broadcast Channels}

\author{Vincent Y. F. Tan,  {\em Senior Member, IEEE}    $\qquad$ Si-Hyeon Lee, {\em Member, IEEE}
\thanks{Vincent~Y.~F. Tan is with the Department of Electrical and Computer Engineering and the Department of Mathematics, National University of Singapore, Singapore 117583 (Email:  vtan@nus.edu.sg). } 
\thanks{Si-Hyeon Lee (Corresponding author) is with the Department of Electrical Engineering, Pohang University of Science and Technology (POSTECH), Pohang, South Korea 37673 (Email: sihyeon@postech.ac.kr).}
  \thanks{The material in this paper was presented in part at IEEE ITW 2018~\cite{TanLee18}. The work of V.~Y.~F.\ Tan was supported in part by a Singapore NRF Fellowship (NRF2017NRF-NRFF001-070 and R-263-000-D02-281) and an NUS Young Investigator Award (R-263-000-B37-133).  The work of  S.-H.\ Lee was supported in part by Basic Science Research Program through the National Research Foundation of Korea (NRF) funded by the Ministry of Education (2017R1D1A1B03034492), and in part by Institute for Information \& communications Technology Promotion (IITP) grant funded by the Korea
government (MSIT) (No. 2018-0-00764, Research on Physical Layer Security with Low Power and Low Latency for Massive IoT  Networks).}
}


\maketitle

\begin{abstract} 
We consider a covert communication scenario where a transmitter wishes to communicate simultaneously to two legitimate receivers while ensuring that the communication is not detected by an adversary, the warden. The legitimate receivers and the adversary observe the transmission from the transmitter via a three-user discrete or Gaussian memoryless broadcast channel. We focus on the case where the ``no-input'' symbol is not redundant, i.e., the output distribution at the warden induced by the no-input symbol is not a mixture of the output distributions induced by other input symbols, so that the covert communication is governed by the square root law, i.e., at most $\Theta(\sqrt{n})$ bits can be transmitted over $n$ channel uses. We show that for such a setting, a simple time-division strategy achieves the optimal throughputs for a non-trivial class of broadcast channels; this is not true for communicating over broadcast channels without the covert communication constraint. Our result implies that a code that uses two separate optimal point-to-point codes each designed for the constituent channels and each used for a fraction of the time is optimal in the sense that it achieves the best constants of the $\sqrt{n}$-scaling for the throughputs. Our proof strategy combines several elements in the network information theory literature, including concave envelope representations of the capacity regions of broadcast channels and El Gamal's outer bound for more capable broadcast channels. 
\end{abstract}  
 
\section{Introduction} \label{sec:intro}
There has been a recent surge of research interest in reliable communications in the presence of an adversary, or a warden, who must be kept incognizant of the presence of communication between the transmitters and receivers. The lack of communication is modelled in discrete channels by sending a specially-designed ``no-input'' symbol $0 \in\calX$ (where $\calX$, a finite set, is the input alphabet of the discrete channel); in Gaussian channels, it is also modelled as sending $0 \in\bbR$. This line of research, known synonymously  as covert communications, communication with low probability of detection (LPD)~\cite{bashgoekeltowsley13,BashNature15,Bash15},   deniability~\cite{chebakshijaggi13,chebakshijaggi14},  or undetectable communication~\cite{LeeBaxley15}, seeks to establish fundamental limits on the throughputs to communicate to the legitimate receiver(s) while ensuring that the signals observed by the warden are statistically close  to the signals if   communication were not present. It was shown by  Bash {\em et al.}~\cite{bashgoekeltowsley13} that in the point-to-point setting, if the legitimate user's channel and the adversary's channel are perfectly known, the number of bits that can be reliably and covertly transmitted over $n$ channel uses scales at most as $\Theta(\sqrt{n})$ for additive white Gaussian  noise  (AWGN) channels. This is colloquially known as the square root law (SRL).  For discrete memoryless channels, the covert communication is also governed by the SRL if the no-input symbol is not redundant, i.e., the output distribution at the warden induced by the no-input symbol is not  a mixture of the output distributions induced by other input symbols. 
Recently, the optimal pre-constant in the $\Theta(\sqrt{n})$ term has also been established  by Bloch~\cite{Bloch16} and Wang, Wornell and Zheng~\cite{WangWornellZheng16}.  

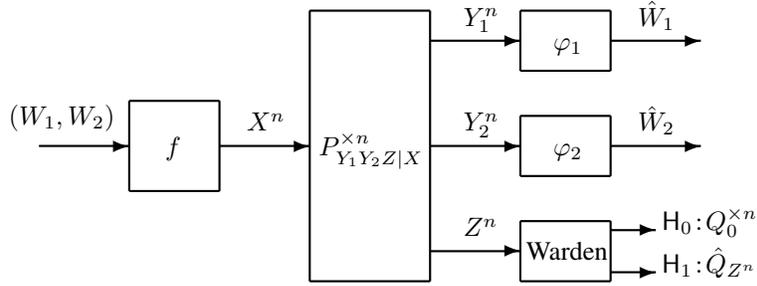
\begin{figure*}[t]
\centering
\setlength{\unitlength}{.4mm}
\begin{picture}(220, 90)
\thicklines
\put(-20, 52){$(W_1,W_2)$}
 \put(-10, 45){\vector(1,0){30}}
\put(20, 30){\line(1, 0){30}}
\put(20, 30){\line(0,1){30}}
\put(50, 60){\line(0,-1){30}}
\put(20, 60){\line(1,0){30}}
\put(50, 45){\vector(1, 0){30}}
\put(80, 0){\line(1, 0){40}}
\put(80, 0){\line(0,1){90}}
\put(120, 0){\line(0,1){90}}
\put(80, 90){\line(1,0){40}}
\put(56, 50){  $X^n$}
\put(32, 42){$f$ } 
 
\put(83, 42){$P_{Y_1 Y_2 Z|X}^{\times n}$} 
\put(128,85){  $Y_1^n$}
\put(128,50){  $Y_2^n$}

\put(128,15){  $Z^n$}

\put(186, 85){  $\hatW_1$}
\put(186, 50){  $\hatW_2$}
\put(120, 80){\vector(1, 0){30}}
\put(180, 80){\vector(1, 0){30}}
\put(120, 45){\vector(1, 0){30}}

\put(120, 10){\vector(1, 0){30}}

\put(180, 17){\vector(1, 0){15}}
\put(194,17){  $\rvH_0 \!:\! Q_0^{\times n}$}
\put(180, 3){\vector(1, 0){15}}
\put(194,3){  $\rvH_1 \!: \!\hatQ_{Z^{n}}$}
\put(150, 35){\line(1, 0){30}}
\put(150, 35){\line(0,1){20}}
\put(180, 35){\line(0,1){20}}
\put(150, 55){\line(1,0){30}}

\put(150, 0){\line(1, 0){30}}
\put(150, 0){\line(0,1){20}}
\put(180, 0){\line(0,1){20}}
\put(150, 20){\line(1,0){30}}

\put(150, 70){\line(1, 0){30}}
\put(150, 70){\line(0,1){20}}
\put(180, 70){\line(0,1){20}}
\put(150, 90){\line(1,0){30}}

\put(180, 45){\vector(1, 0){30}}

\put(161, 77){$\varphi_1$ } 
\put(161, 42){$\varphi_2$ } 
\put(152,7){Warden}
  \end{picture}
  \caption{Illustration of two-user BC with a warden. The key which is accessible to $f, \varphi_1$, and $\varphi_2$ is not shown here.}
  \label{fig:model}
\end{figure*}

In this paper, we extend the above model and results to a multi-user  scenario~\cite{elgamal} in which there is one transmitter, two legitimate receivers and, as usual, one warden. See Fig.~\ref{fig:model}. We are interested in communicating reliably and simultaneously to the two receivers over the same medium while ensuring that the warden remains incognizant of the presence of any communication. We call our model a {\em two-user discrete memoryless broadcast channel (BC) with a warden}. This communication model  mimics the scenario of a military general  delivering commands to her/his multiple subordinates while, at the same time, ensuring that the probability of the communication being detected by a furtive enemy, the warden, is vanishingly small. Note that the enemy is not interested in the precise commands {\em per se} but on whether or not communication between the general and her/his subordinates is actually happening in order to pre-empt a possible attack. We establish the fundamental   limits for communicating in this scenario when the no-input symbol is not redundant. Somewhat surprisingly, we show that the most basic multi-user communication scheme of time-division~\cite[Sec.~5.2]{elgamal}  is optimal for a  class of BCs. This implies that a code designed for such BCs that uses two {\em separate} point-to-point codes, each designed for the constituent channels and each used for a fraction of the time (blocklength) is optimal  in the sense that it achieves the best constants of the $\sqrt{n}$-scaling of the throughput. We would like to emphasize that time-division is not optimal for the vast majority of BCs in the absence of the covert communication constraint~\cite[Chapters~5 \&~8]{elgamal}. In fact, the set of BCs for which time-division is optimal without the covert communication constraint has measure zero but with the covert communication constraint, this measure is not zero (see Fig.~\ref{fig:con}).

\subsection{Related Work} \label{sec:related} 
Covert communication is related with   steganography in the sense that both aim to hide information against a warden. In many instances of steganography, the SRL has been observed \cite{Anderson96, KerPevny08,Korzhik,FillerFridrich09,Ker09,Ker:2013:MSS:2482513.2482965}.  In steganography, a cover text is given to the encoder and a message is embedded into the cover text when the communication is active.  The message-embedded text is called the  stegotext  and the warden should not be able to distinguish the stegotext from the cover text. In particular, a setup closely related to our problem is the scenario in which  the stegotext is generated through a memoryless channel as in~\cite{Korzhik, FillerFridrich09,Ker09}. It is interesting to observe that  for steganography, a source sequence (cover text) is given to the encoder and the encoder controls the channel (regarding the cover text as the input and the stegotext as the output), while in covert communications, the encoder generates a source sequence (codeword) for each message and then the source sequence is sent through a given communication channel. Since the given condition and the control function are swapped, the two classes of problems require different analyses. For example, the positive steganography rate shown in \cite{CraverYu} relys on the fact that the encoder can modify some part of the covertext, while such a technique is not applicable in standard communication systems unless the transmitter has access to  the warden's noise.



This paper focuses on the information-theoretic aspects of covert communications for which there has been a flurry of recent work. 
In particular, refined asymptotics on the fundamental limits of covert communications over memoryless channels from the second-order~\cite{Mehrdad16} and error exponent~\cite{Mehrdad17} perspectives have been studied. In addition,  the fundamental limits of covert communications for  channels with random state known at the transmitter~\cite{LeeState17} and classical-quantum channels~\cite{Wang16} have also been established. In contrast,  work on multi-user extensions of covert communications is relatively sparse. Of note are the works by Arumugam and Bloch~\cite{Arumugam16,Arumugam17}. In~\cite{Arumugam16}, the authors derived the fundamental limits of covert communications over a multiple-access channel (MAC). The authors showed that  if the MAC to the legitimate receiver is ``better'', in a precise sense, than the one to the adversary,  then the legitimate users can reliably communicate on the order of $\Theta(\sqrt{n})$ bits per $n$ channel uses with arbitrarily LPD  without using a secret key. The authors also quantified the pre-constant terms exactly.  In~\cite{Arumugam17}, the authors considered a BC  communication model. However, note that the model in~\cite{Arumugam17} is significantly different from that in the present work. In~\cite{Arumugam17}, the authors were interested in transmitting two messages, one common and another covert, over a BC where there are {\em two} receivers, one legitimate and the other the warden. The common message is to be communicated to both parties, while the covert message has LPD from the perspective of the warden. This models the scenario of embedding covert messages in an innocuous codebook and generalizes existing works on covert communications in which innocent behavior corresponds to lack of communication between the legitimate receivers. In our work, there are {\em three} receivers, two legitimate and the other denotes the warden. Both messages that are communicated can be considered as covert message from the perspective of the warden. See Fig.~\ref{fig:model} for our model.

\subsection{Main Contribution}
Our main contribution is to establish the covert capacity region (the set of all achievable pre-constants of the throughputs which scale as $\Theta(\sqrt{n})$) of some two-user  memoryless BCs with a warden when the no-input symbol is not redundant. While the (usual) capacity region of the two-user discrete memoryless BC  is a long-standing open problem in network information theory, we show that under the covert communications constraint, the covert capacity region admits a particularly simple expression for a class of BCs we explicitly  identify (see Condition~\ref{cond1}). This region implies  that time-division transmission is optimal for this class of two-user BCs. We emphasize the BC  does not have to be degraded, less noisy or more capable~\cite[Chapter~5]{elgamal}.  Our main result is somewhat analogous to that of Lapidoth, Telatar and Urbanke~\cite{LTU03} who showed that time-division is optimal for  wide-band broadcast communication over Gaussian, Poisson, ``very noisy'' channels, and average-power limited fading channels. However, the analysis in~\cite{LTU03} is restricted to stochastically degraded BCs.

\begin{figure*}[t]
\centering
\setlength{\unitlength}{.4mm}
\begin{picture}(220, 90)
\multiput(20,0)(10,0){21}{\line(0,1){15}}
\multiput(20,30)(10,0){21}{\line(0,1){15}}
\multiput(20,60)(10,0){21}{\line(0,1){15}}

\thicklines
\put(20, 0){\line(1,0){200}}
\put(20, 15){\line(1,0){200}}
\put(20, 0){\line(0,1){15}}
\put(220, 0){\line(0,1){15}}
\put(200, 0){\line(0,1){15}}
\put(-23, 5){$x^n(w_1,w_2)$}
\put(60,0){\textcolor{black}{\rule{10\unitlength}{15\unitlength}}}
\put(200,0){\textcolor{black}{\rule{10\unitlength}{15\unitlength}}}
\put(100,0){\textcolor{black}{\rule{10\unitlength}{15\unitlength}}}

\put(20, 30){\line(1,0){200}}
\put(20, 45){\line(1,0){200}}
\put(20, 30){\line(0,1){15}}
\put(220, 30){\line(0,1){15}}

\put(-13, 35){$u_2^n(w_2)$}
\put(110, 20){$| \,|$}
\put(60,30){\textcolor{black}{\rule{10\unitlength}{15\unitlength}}}
\put(200,30){\textcolor{black}{\rule{10\unitlength}{15\unitlength}}}

\put(20, 60){\line(1,0){200}}
\put(20, 75){\line(1,0){200}}
\put(20, 60){\line(0,1){15}}
\put(220, 60){\line(0,1){15}}

\put(-13, 65){$u_1^n(w_1)$}
\put(110, 50){$\oplus$}

\put(0, 20){$| \,|$}
\put(0, 50){$\oplus$}
\put(100,60){\textcolor{black}{\rule{10\unitlength}{15\unitlength}}}
\end{picture}
\caption{Illustration of intuition of optimality of time-division. Shaded (resp.\ unshaded) boxes indicate that that coordinate equals $1$ (resp.\ $0)$. Because $u_1^n(w_1)$ and $u_2^n(w_2)$ are very sparse, the set of locations of the $1$'s in $x^n(w_1,w_2)$ is highly likely to be a {\em disjoint} union of those in $u_1^n(w_1)$ and $u_2^n(w_2)$.} 
\label{fig:time-share}
\end{figure*}
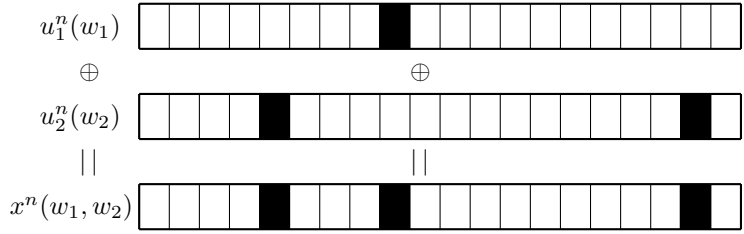

The first analytical tool that is used to prove our main theorem is a converse bound derived by El Gamal~\cite{elgamal79} initially designed for more capable BCs. However, it turns out to be  a  useful starting point for our setting.  We manipulate this bound into a form that is reminiscent of an outer bound for degraded BCs. Subsequently, we express the $\lambda$-sum throughput of this outer bound, for appropriately chosen $\lambda\ge 1$, in terms of upper concave envelopes~\cite{Nair2013}. This circumvents the need to identify the optimal auxiliary random variable for specific BCs. The identification of optimal auxiliaries is typically possible only if the BC  possesses some special structure. For example for the binary symmetric BC, Mrs.\ Gerber's lemma~\cite{WZ73} is a key ingredient in simplifying the converse bound. Similarly for the Gaussian BC, a highly non-trivial result known as the entropy power inequality~\cite{Shannon48,StamEPI59} is needed to obtain an explicit expression for the capacity region. Note that both these channels are degraded.  Our main results applies to some classes of two-user discrete memoryless BCs (that do not subsume degraded BCs nor is subsumed by degraded BCs). We employ concave envelopes and take into consideration that under the covertness constraint, the weight of the codeword is necessarily  vanishingly small~\cite{WangWornellZheng16,Bloch16}. Augmenting some basic analytical arguments (e.g., \cite[Lemma~1]{Bloch16}) to these existing techniques  allows us to  simplify the outer bound to conclude that the time-division inner bound~\cite[Sec.~5.2]{elgamal} is optimal for a class of BCs.

\subsection{Intuition Behind The Main Result} \label{sec:intuition}
We now provide   intuition as to why time-division transmission is optimal for some BCs with covertness constraints. Since symmetric channels satisfy our condition (point~3  in the discussion following Condition~\ref{cond1}), we use the binary symmetric BC~\cite[Example~5.3]{elgamal}, degraded in favor of $Y_1$ (say), as an example to illustrate this. Let $Y_j = X\oplus\Psi_j$ where $\Psi_j\sim\mathrm{Bern}(p_j)$ for $j = 1,2$, i.e., the channel from $X$ to $Y_j$ is a binary symmetric channel (BSC). It is well known that in the absence of the covertness constraint, superposition coding~\cite[Sec.~5.3]{elgamal} \cite{cover72}  is optimal,  and time-division transmission is strictly suboptimal (unless the transmitter communicates with only a single receiver).  The coding scheme is illustrated in Fig.~\ref{fig:time-share} where  $u_1^n(w_1),u_2^n(w_2)\in\bbF_2^n$ are generated independently, $u_2^n(w_2)$ denotes the cloud center and $x^n(w_1,w_2)= u_1^n(w_1)\oplus u_2^n(w_2)$ denotes the satellite codeword. Because in covert communications, $x^n(w_1,w_2)$ is required to have low (Hamming) weight~\cite{Bloch16,WangWornellZheng16}, either both $u_1^n(w_1)$ and $u_2^n(w_2)$  have low weight or both   have high weight. Let us assume it is the former without loss of generality. Then, the set of locations of the $1$'s in the satellite codeword is likely to be a {\em disjoint} union of the sets of $1$'s in  $u_1^n(w_1)$ and $u_2^n(w_2)$. Thus the relative weight of $x^n(w_1,w_2)$, denoted as $\alpha_n$, can be decomposed into the relative weights of $u_1^n(w_1)$ and $u_2^n(m_2)$, denoted as $\rho\alpha_n$ and $(1-\rho)\alpha_n$  respectively for some $\rho \in [0,1]$. Thus, the superposition coding inner bound for the degraded BC~\cite[Theorem~5.2]{elgamal} \cite{cover72} with $X=U_1\oplus U_2$  reads 
\begin{align}
R_1 &\le I(X;Y_1|U_2)= I(U_1;U_1\oplus \Psi_1)\approx \rho\alpha_n L_1^*, \label{eqn:R1} \\
 R_2&\le I(U_2;Y_2)= I(U_2;U_2\oplus \tilde{\Psi}_2 )  \approx  (1- \rho) \alpha_n L_2^*,\label{eqn:R2}
\end{align}
 where $L_1^*$ and $L_2^*$ are the ``covert capacities''~\cite{WangWornellZheng16,Bloch16} of the constituent point-to-point binary symmetric channels (see Theorem~\ref{def:cov_cap}) and $\tilde{\Psi}_2:=U_1\oplus \Psi_2$ is a   Bernoulli random variable with parameter $\rho\alpha_n\ast p_2 \approx p_2$ due to the low weight constraint on $U_1$ (its relative weight is $\rho\alpha_n$). Thus,~\eqref{eqn:R1} and~\eqref{eqn:R2}  suggest  that time-division is optimal, at least for symmetric BCs but we show the same is true for a larger class of BCs. 

\subsection{Paper Outline}
This paper is structured as follows. In Section~\ref{sec:formulation}, we formulate the problem and define relevant quantities of interest precisely. In Section~\ref{sec:main_res}, we state our assumptions and main results. We also  provide some qualitative interpretations. The main result, corollaries, and bounds on the required length of the secret key are proved in Sections~\ref{sec:prf_thmmain},~\ref{sec:prf_coro},  and~\ref{sec:prf_key} respectively. We conclude our discussion in Section~\ref{sec:concl} where we also suggest avenues for future research.
\subsection{Notation}
We adopt standard information-theoretic notation, following the text by El Gamal and Kim~\cite{elgamal} and on occasion, the book by Csisz\'ar and K\"orner~\cite{Csi97}. We use $h_{\rmb}(q):=-q\log q-(1-q)\log(1-q)$ to denote the binary entropy function. $D(P\|Q):=\sum_z P(z)\log \frac{P(z)}{Q(z)}$ and $\bbV(P,Q):=\frac{1}{2}\sum_z |P(z)-Q(z)|$ are  the relative entropy and the total variation distance respectively. We use $I(X;Y)$ and $I(P,W)$ to denote the mutual information of $(X,Y)\sim P\times W$. Throughout $\log $ is taken to an arbitrary base. We also use $h(\cdot)$ to denote the differential entropy. The notation $a\ast b:=a(1-b)+b(1-a)$ denotes the binary convolution operator. For two discrete distributions, $P$ and $Q$ defined on the same alphabet $\calX$, we say that $P$ is {\em absolutely continuous with respect to $Q$}, denoted as $P\ll Q$ if  for all $x\in\calX$, $Q(x)=0$ implies that $P(x)=0$.  Other notation will be introduced as needed in the sequel. 

\section{Problem Formulation}\label{sec:formulation}
A  {\em  discrete memoryless\footnote{We omit the qualifier ``discrete memoryless'' for brevity in the sequel. We will mostly discuss discrete memoryless BCs in this paper. However, in Corollary~\ref{cor:gauss}, we present results for the BC with a warden when the constituent channels are Gaussian.}  two-user BC  with a warden}  $(\calX,\calY_1,\calY_2, \calZ,P_{Y_1,Y_2,Z|X})$ consists of a    channel input alphabet $\calX$, three    channel output alphabets $\calY_1$, $\calY_2$, and $\calZ$, and a   transition matrix $P_{Y_1, Y_2,Z |X}$.  The output alphabets $\calY_1$ and $\calY_2$ correspond to the two legitimate receivers and $\calZ$ corresponds to that of the warden. 
Without loss of generality, we let $\calX=\{0, 1, \ldots, K\}$.  We let $0\in\calX$ be the ``no input'' symbol that is sent when no communication takes place and define $Q_x=P_{Z|X}(\cdot|x)$ for each $x\in\calX$.  The BC is used $n$ times in a memoryless manner. If no communication takes place, the warden at receiver $\calZ$ observes $Z^n$,  which is distributed according to $Q_0^{\times n}$, the $n$-fold product distribution of $Q_0$.  If communication occurs, the warden observes $\hatQ_{Z^n}$, the output distribution induced by the code. 
  For convenience, in the sequel, we often denote the two marginal channels corresponding to the  two legitimate receivers as $W=P_{Y_1|X}$ and $V=P_{Y_2|X}$ respectively.  
  
 The transmitter and the receivers are assumed to share a secret key $S$ uniformly distributed over a set $\calK$.  We will, for the most part, assume that the key is sufficiently long, i.e., the set $\calK$ is sufficiently large. However, we will bound the length of the key in Section~\ref{sec:key}.  The transmitter and the receiver aim to  construct a code that is both reliable and covert. Let the  messages to be sent  be $ W_1$ and $W_2$. These messages are assumed to be independent and also independent of $S$. Also let their reconstructions at the receiver $\calY_j$ be $\hatW_j$ for $j = 1,2$.  As usual, a  code is said to be {\em  reliable} if the probability of error $\Pr(\cup_{j=1}^2\{\hatW_j\ne W_j\})$ vanishes as $n \to\infty$. The  code is {\em covert}  if it
is difficult  for the warden to determine whether the transmitter is sending a message (hypothesis $\rvH_1 $) or not
(hypothesis $\rvH_0$). Let $\pi_{1|0}$ and $\pi_{0|1}$ denote the probabilities of false alarm (accepting $\rvH_1 $ when the transmitter is
not sending a message) and missed detection (accepting $\rvH_0$ when the transmitter is sending a message),
respectively. Note that a blind test (one with no side information) satisfies $\pi_{1|0}+\pi_{0|1}= 1$. 
  The warden's optimal hypothesis test satisfies 
\begin{align}
\pi_{1|0}+\pi_{0|1} &=1- \bbV(  \hatQ_{Z^n} , Q_0^{\times n}  ) \label{eqn:var}\\
& \ge 1-\sqrt{D(\hatQ_{Z^n}\| Q_0^{\times n}  )}  , \label{eqn:pinsker}
\end{align}
where \eqref{eqn:pinsker} follows by Pinsker's inequality; 
see~\cite{LehmannRomano:TSH,Bloch16}. Hence, covertness is guaranteed if the  relative entropy  between the observed distribution $\hatQ_{Z^n}$  and the product  of no communication distribution $Q_0^{\times n}$ is bounded by a  small $\delta>0$. 

Note that if $\supp(P_{Z|X}(\cdot|x))\nsubseteq \supp(Q_0)$ for some $x\in \calX$, such $x$ should not be transmitted, otherwise it is not possible for $D(\hatQ_{Z^n}\| Q_0^{\times n}  )$ to vanish~\cite{WangWornellZheng16}. Hence, by dropping all such input symbols as well as all output symbols not included in $\supp(Q_0)$, we assume throughout that $\supp(Q_0)=\calZ$.  In addition, we assume that the no-input symbol $0$ is not redundant  i.e.,  $P_{Z|X}(\cdot|0)  \notin \mathrm{conv}\{P_{Z|X}(\cdot|x')\colon x'\in \mathcal{X}, x'\neq 0\}$ where $\mathrm{conv}\{\cdot\}$ denotes the convex hull. If the symbol $0$ is redundant, there exists a sequence of codes for which $D(\hatQ_{Z^n}\| Q_0^{\times n}  )=0$ for all $n$~\cite{WangWornellZheng16} so $\pi_{1|0}+\pi_{0|1}= 1$  (i.e., the warden's test is always blind) and transmitting at positive rates is possible; this is a regime we do not consider in this paper.

A code for the BC with a covertness constraint and the covert capacity region are defined formally as follows.
\begin{definition} \label{def:code}
An {\em $(n,M_{1n}, M_{2n}, K_n, \eps,\delta )$-code} for the BC  with a warden $(\calX,\calY_1,\calY_2,\calZ, P_{Y_1,Y_2,Z|X})$  and with  a covertness constraint  consists of
\begin{itemize}
\item Two message sets $\calM_j := \{1,\ldots, M_{jn}\}$ for $j = 1,2$;
\item Two independent messages uniformly distributed over their respective message sets, i.e., $W_j \sim\mathrm{Unif}(\calM_j)$ for $j = 1,2$;
\item One secret key set $\calK :=\{1,\ldots, K_n\}$; 
\item One encoder $f: \calM_{1n}\times\calM_{2n} \times\calK\to\calX^n$;
\item Two decoders $\varphi_j :\calY_j^n\times\calK\to\calM_j$ for $j = 1,2$;
\end{itemize}
such that the following constraints hold:
\begin{align}
  \Pr\big(\cup_{j=1}^2\{\hatW_j\ne W_j\}  \big) &\le\eps ,\label{eqn:rel_constrain}\\
D (\hatQ_{Z^n} \|  Q_0^{\times n }  )&\le\delta. \label{eqn:convert_constraint}
\end{align}
\end{definition}
For most of our discussion,  we ignore the secret key set (i.e., we assume that the secret key is sufficiently long) for the sake of simplicity and refer to the family of codes above with secret key sets  of arbitrary sizes  as $(n,M_{1n}, M_{2n},  \eps,\delta )$-codes. We will revisit the effect of the key size in Section~\ref{sec:key}. 

\begin{definition} \label{def:region}  
We say that the pair $(L_1, L_2)\in\bbR_+^2$ is {\em $(\eps,\delta)$-achievable} for the  BC   with a warden and with a covertness constraint if there exists a sequence of $(n,M_{1n}, M_{2n}, \eps_n,\delta)$-codes such that
\begin{align}
\liminf_{n\to\infty}\frac{1}{\sqrt{n\delta}}\log M_{jn} &\ge L_j ,\quad j \in\{1,2\} ,\label{eqn:rate_cons}\\
\limsup_{n\to\infty} \eps_n &\le\eps. \label{eqn:rel_constr}
\end{align}
Define the {\em $(\eps,\delta)$-covert capacity region} $\calL_{\eps,\delta} \subset\bbR_+^2$ to be  the closure of all $(\eps,\delta)$-achievable pairs of $(L_1, L_2)$. We are interested in  the {\em $\delta$-covert capacity region} 
\begin{equation}
\calL_\delta := \bigcap_{\eps \in (0,1)} \calL_{\eps,\delta} =\lim_{\eps\to 0}\calL_{\eps,\delta}.
\end{equation}
\end{definition}
Note that $L_1$ and $L_2$  in~\eqref{eqn:rate_cons} are measured in bits (or information units) per square root channel use. 

%

We will also need the notion of covert capacities  for point-to-point discrete memoryless channels (DMCs) with a warden  $(\calX,\calY,\calZ, P_{Y,Z|X})$~\cite{WangWornellZheng16,Bloch16}. This scenario corresponds to the above definitions with $Y_1=Y$ and $Y_2=\emptyset$.  Recall that the {\em  chi-squared distance} between two distributions $Q_0$ and $Q_1$ supported on the same alphabet $\calZ$ is defined as
\begin{align}
\chi_2(Q_1\|Q_0)&:=\sum_{z\in\calZ} \frac{(Q_1(z)-Q_0(z))^2}{Q_0(z)}.
\end{align}

 \begin{theorem}[Bloch~\cite{Bloch16} and Wang, Wornell, Zheng~\cite{WangWornellZheng16}]\label{def:cov_cap} 
Let  $(\calX,\calY,\calZ, P_{Y,Z|X})$ be a DMC with a warden in which $W:=P_{Y|X}$ and $Q_k := P_{Z|X}(\cdot |k)$ for $k\in\calX$. We assume that it satisfies  $W(\cdot|k)\ll W(\cdot|0)$ for all $k\in\calX\setminus\{0\}$, $\mathrm{supp}(Q_0)=\calZ$ and $0\in\calX$ is not redundant. Then its covert capacity  is 
 \begin{equation}
 L^*(P_{Y,Z|X}) :=\max_{\mathbf{p} }\sqrt{\frac{2 \big(\sum_{k=1}^Kp_k D(W(\cdot |k ) \| W(\cdot | 0 )) \big)^2}{\chi_2\big(\sum_{k=1}^K p_k Q_k\big\|Q_0\big)}} \label{eqn:Lgen}
 \end{equation}
 where the maximization   extends over all length-$K$ probability vectors $\bp$ (i.e., $\bp=[p_1,\ldots,p_K]^T$ where $p_k\ge 0$ and $\sum_{k=1}^K p_k=1$). If $\calX = \{0,1\}$, i.e., $P_{Y,Z|X}$ has a binary input, then  the maximization over $\bp$ in~\eqref{eqn:Lgen} is unnecessary and
 \begin{equation}
 L^*(P_{Y,Z|X}) := \sqrt{ \frac{2 D(W(\cdot|1) \| W(\cdot|0))^2}{\chi_2(Q_1\|Q_0 )}}. \label{eqn:Lbin}
 \end{equation}
 \end{theorem}
As previously mentioned, we assume that $Q_1\ll Q_0$ in the binary input case  (or more generally, $\sum_{k=1}^K p_k^* Q_k\ll Q_0$ where $\bp^*=[p_1^*,\ldots, p_K^*]^T$ is any maximizer of~\eqref{eqn:Lgen}). Otherwise, covert communication is impossible~\cite[Appendix~G]{Bloch16}. With this assumption  and the fact that $0\in\calX$ is not redundant, $L^*(P_{Y,Z|X})$ as defined in~\eqref{eqn:Lgen} and~\eqref{eqn:Lbin} is finite.
\section{Main Results} \label{sec:main_res}
In this section, we present our main results. In Section~\ref{sec:cond}, we state a   condition on BCs, which we call Condition~\ref{cond1}.  In Section~\ref{sec:memoryless}, we state our main result assuming the BC satisfies Condition~\ref{cond1} and the key is of sufficiently long length. We interpret our result in Section~\ref{sec:remarks} by placing it in context via several remarks. In Section~\ref{sec:degraded}, we specialize our   result to two degraded BCs and show that standard techniques apply for such   models. In Section~\ref{sec:key}, we extend our main result to the case where the key size is also a parameter of interest.
\subsection{A Condition on BCs} \label{sec:cond}
In the following, we consider the following condition on BCs that allows us to show that time-division is optimal.
\begin{condition} \label{cond1}
Fix a BC with a  warden $P_{Y_1,Y_2,Z|X}$. Let the   covert capacities of $P_{Y_1,Z|X}$ and $P_{Y_2,Z|X}$ be $L_1^* := L^*(P_{Y_1,Z|X})$ and  $L_2^* := L^*(P_{Y_2,Z|X})$ respectively. 
\begin{itemize} 
\item If $L_1^*\ge L_2^*$,  we assume that 
\begin{equation}
\max_{P_X} \frac{I(X;Y_1)}{I(X;Y_2)}\le \frac{L_1^*}{L_2^*}\label{eqn:cond1}
\end{equation}
\item Otherwise if $L_2^*\ge L_1^*$, we assume that 
\begin{equation}
\max_{P_X} \frac{I(X;Y_2)}{I(X;Y_1)}\le  \frac{L_2^*}{L_1^*}\label{eqn:cond2}
\end{equation}
\end{itemize}
\end{condition}
A few remarks concerning Condition~\ref{cond1} are in order. 
 
 \begin{enumerate}[leftmargin=*]
\item     Condition \ref{cond1}, which is easy to check numerically as the optimizations over $P_X$ are over compact sets, neither subsumes nor is subsumed by degradedness or any other ordering of $W=P_{Y_1|X}$ and $V=P_{Y_2|X}$. That is, we can show that there exists some degraded BCs that do not satisfy Condition~\ref{cond1} and there are also non-degraded BCs that satisfy Condition~\ref{cond1}. 
\item Condition~\ref{cond1} is significantly simplified in the binary-input case. Let $W=P_{Y_1|X}$, $V=P_{Y_2|X}$, and  $W_\gamma(y) :=\sum_{x\in\calX} P_\gamma(x) W(y|x),y\in\calY$ where  $P_\gamma$ is the Bernoulli distribution with probability of $1$ being $\gamma \in [0,1]$, i.e., 
\begin{equation}
P_{\gamma} (x) := \left\{ \begin{array}{cc}
1-\gamma& x = 0 \\
 \gamma & x = 1 \\
\end{array} \right. . \label{eqn:Palpha_n}
\end{equation} 
Similarly define $V_\gamma$.  Then, one can use~\eqref{eqn:Lbin} and Lemma~\ref{lemma:weight}  (to follow)   to show that~\eqref{eqn:cond1} is equivalent to 
\begin{equation}
 \frac{D(W_1 \| W_0 )}{D(V_1 \| V_0 )} \le \min_{\gamma\in [0,1]}  \frac{D(W_\gamma \| W_0) }{D(V_\gamma \| V_0 )}. \label{eqn:cond_bin}
\end{equation}
Thus the verification  of Condition~\ref{cond1}  for binary-input BCs reduces to a  line search over $[0,1]$.
\begin{figure}
\centering
\includegraphics[width=1.01\columnwidth]{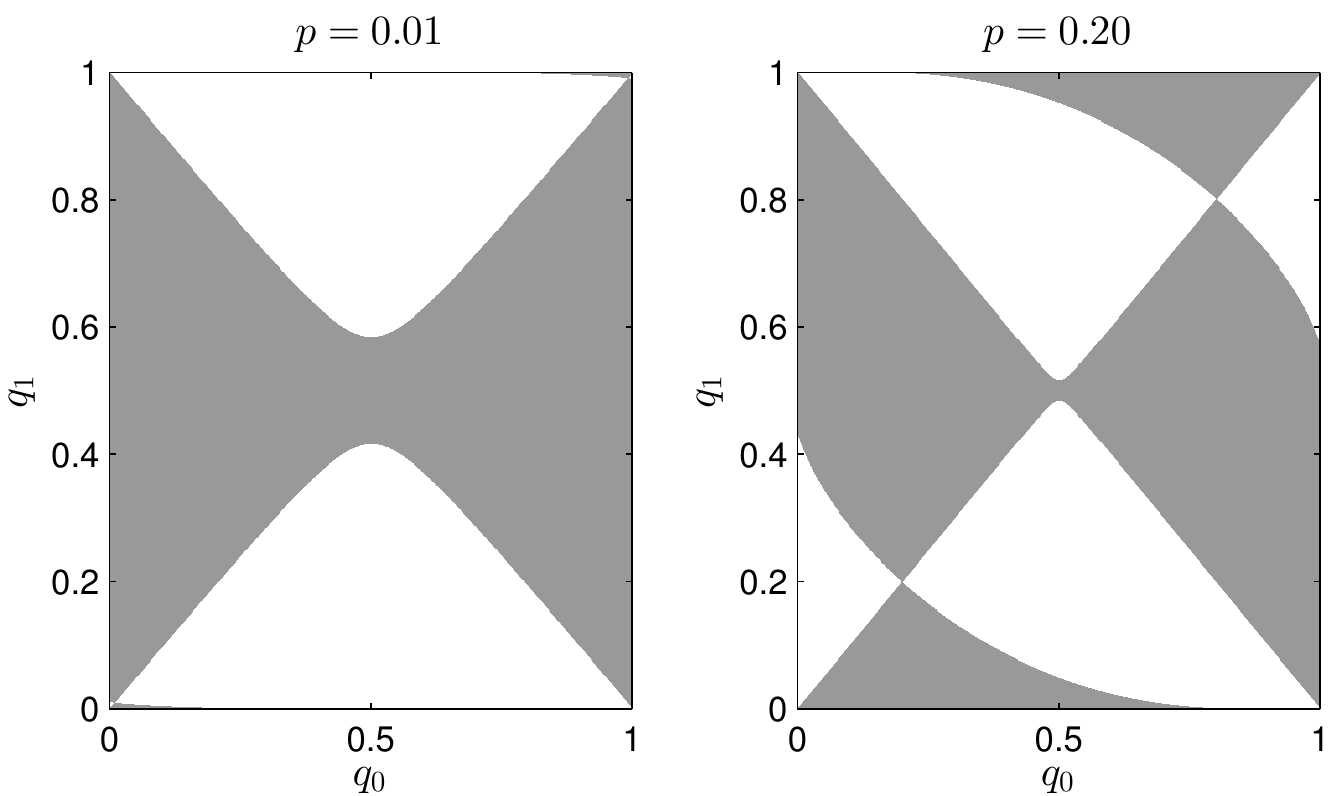}
\caption{The set of values of $(q_0,q_1) \in [0,1]^2$ (parameters of the matrix $V$ in~\eqref{eqn:Vmat})  such that   Condition~\ref{cond1} is  satisfied (resp.\ not satisfied) is indicated in gray (resp.\ white).}
\label{fig:con}
\end{figure}

\item To illustrate this condition, we consider the scenario in which\footnote{We use the notation $W=\mathrm{BSC}(q)$ if it is a binary-input, binary-output channel in which $Y=X\oplus \Psi$ where $\Psi\sim\mathrm{Bern}(q)$.} $W=\mathrm{BSC}(p)$ where $p\in\{0.01,0.20\}$ and $V$ is a (generally) asymmetric binary-input, binary-output channel such that the transition matrix reads
\begin{equation}
V = \begin{bmatrix}
1-q_0  & q_0\\
q_1  & 1-q_1
\end{bmatrix}, \label{eqn:Vmat}
\end{equation}
for $q_0,q_1\in [0,1]$. In Fig.~\ref{fig:con}, we show the range of values of $(q_0,q_1) \in [0,1]^2$ such that   Condition~\ref{cond1} is satisfied (indicated in gray). Also we note  that the ``diagonal'' values of $(q_0,q_1)$ in which $q_0=q_1$   satisfy Condition~\ref{cond1}. Hence,  if $V$ is a BSC,  Condition~\ref{cond1} is satisfied. More generally, we can verify numerically that if $W$ and $V$ are both BSCs, then Condition~\ref{cond1}, or equivalently~\eqref{eqn:cond_bin} for the case $L_1^*\ge L_2^*$, is satisfied. 
\end{enumerate}

\subsection{Time-Division is Optimal for Some BCs} \label{sec:memoryless}
Our main result is a complete characterization of the $\delta$-covert capacity region for all   BCs satisfying Condition~\ref{cond1} and  certain absolute continuity conditions. 
\begin{theorem} \label{thm:main}
Assume that a BC  with a warden $P_{Y_1,Y_2,Z|X}$  is such that Condition~\ref{cond1} is satisfied and  the constituent DMCs $W:=P_{Y_1|X}$ and $V:=P_{Y_2|X}$  satisfy $W(\cdot|k)\ll W(\cdot|0)$ and $V(\cdot|k)\ll V(\cdot|0)$ for all $k\in\calX\setminus\{0\}$.   Also assume that the length of the secret  key is sufficiently large. 
Then, for all $\delta> 0$, the 
$\delta$-covert capacity region is 
\begin{equation}
\calL_\delta=  \left\{ (L_1, L_2)\in\bbR_+^2: \frac{L_1}{L_1^*}+\frac{L_2}{L_2^* }\le 1  \right\} .  \label{eqn:covert_cap_reg}
\end{equation}
\end{theorem}
Theorem \ref{thm:main} is proved in Section~\ref{sec:prf_thmmain}.  
\subsection{Remarks on the Main Theorem} \label{sec:remarks}
A few remarks are in order. 

\begin{enumerate}[leftmargin=*]

 \item First, note that~\eqref{eqn:covert_cap_reg} implies that under the covert communication constraints, time-division transmission is optimal for all  BCs satisfying Condition~\ref{cond1}.  The achievability part  simply involves  two optimal covert communication codes, one for each DMC with a warden. The first code, designed for $P_{Y_1,Z|X}$, is employed over $\lfloor\rho n\rfloor$  channel uses where $\rho\in [0,1]$. The second code, designed for $P_{Y_2,Z|X}$, is employed over the remaining $n- \lfloor\rho n\rfloor$ channel uses. However, because the normalization of $\log M_{jn}$ is $\sqrt{n\delta}$, we need a slightly more subtle time-division argument. To do so,    fix $\delta' < \delta$, then from Theorem \ref{def:cov_cap}, we know that there exists a sequence of codes that allows transmission of
\begin{equation}
\log M_{1n} \cong \sqrt{\rho n\delta'}L_1^*
\end{equation}
bits for user $1$ over  $\lfloor\rho n\rfloor$  channel uses  and with covertness constraint  (upper bound of the divergence in~\eqref{eqn:convert_constraint}) $\delta'$ and 
\begin{equation}
 \log M_{2n}\cong \sqrt{(1-\rho)n(\delta-\delta')}L_2^*
\end{equation}
bits for user $2$ over $n- \lfloor\rho n\rfloor$  channel uses and with covertness constraint $\delta-\delta'$.  Choosing $\delta'=\rho\delta$ and combining these two codes, 
achieves the covertness constraint $\delta$ and the rate point $(\rho L_1^*, (1-\rho)L_2^*)$ which is on the boundary of $\calL_\delta$. By varying $\rho\in [0,1]$, we achieve the whole boundary and hence the entire region in~\eqref{eqn:covert_cap_reg}. 
 
 Note that time-division is strictly suboptimal for the vast majority of BCs in the absence of the covert communication constraint. Indeed, one needs to perform superposition coding~\cite{cover72} to achieve all points in the capacity region for degraded, less noisy and more capable BCs.  Thus, the covert communication constraint significantly simplifies the optimal coding scheme for BCs satisfying Condition~\ref{cond1}.

 \item  The converse of Theorem~\ref{thm:main} thus constitutes the main contribution of this paper. To obtain an explicit outer bound for the capacity region  for BCs that satisfy some ordering---such as degraded, less noisy or more capable BCs~\cite[Chapter~5]{elgamal}---one often has to resort to the identification of the optimal auxiliary random variable-channel input pair $(U,X)$ in the  capacity region of these classes of BCs. However, this is only possible for specific BCs;
see Corollaries~\ref{cor:bsc} and~\ref{cor:gauss} to follow. For general (or even arbitrary degraded) BCs, this is, in general, not possible. Our workaround involves first starting with an  outer bound of  the capacity region for general memoryless BCs by El Gamal~\cite{elgamal79}.  We combine the inequalities in the  outer bounds and use this to upper bound  a linear combination of the two throughputs  in terms  of the concave envelope  of a linear combination of mutual information terms~\cite{Nair2013}. This allows us to circumvent  the need to explicitly characterize $(U,X)$ since $U$ is no longer present in this concave envelope characterization. We then exploit Condition~\ref{cond1} and some approximations to obtain the desired  the outer bound to~\eqref{eqn:covert_cap_reg}. The appeal of this approach is not only that we do not need to find the optimal $(U,X)$. 



 \item Let us say that the {\em absolute continuity condition} holds for $W$ if $W(\cdot|k)\ll W(\cdot|0)$ for all $k\in\calX\setminus\{0\}$. Then Bloch~\cite[Appendix~G]{Bloch16}  showed that if $W$ does not satisfy the absolute continuity condition, $\Theta(\sqrt{n}\log n)$ bits per $n$ channel uses can be covertly transmitted. In our setting, the same is true for the constituent channels; if $W$ satisfies the absolute continuity condition but $V$ does not, $\Theta(\sqrt{n})$ bits per $n$ channel uses  can be transmitted to $Y_1$, while $\Theta(\sqrt{n}\log n)$ bits per $n$ channel uses can be transmitted to~$Y_2$. 

 \item Finally, suppose that the covertness condition in~\eqref{eqn:convert_constraint} is replaced by a total variation constraint of the form 
\begin{equation}
\bbV( \hatQ_{Z^n}, Q_0^{\times n}  )\le\delta,\quad\mbox{for some}\quad \delta\in (0,1). 
\end{equation}
Then, by using the techniques to prove~\cite[Theorem~2]{TahBloch17},  one can easily see that the covert capacity region is   the same as \eqref{eqn:covert_cap_reg} in Theorem~\ref{thm:main} except that in the formulation,~\eqref{eqn:rate_cons} is replaced by 
\begin{equation}
\liminf_{n\to\infty}\frac{1}{\sqrt{n\Gamma_\delta}}\log M_{jn} \ge L_j ,\quad j \in\{1,2\} ,\label{eqn:rate_cons_var}
\end{equation}
where $\Gamma_\delta:=\sqrt{2}\rmQ^{-1}(\frac{1-\delta}{2})$ and $\rmQ(\cdot)$ is the complementary cumulative distribution function of a standard Gaussian. Hence, only the normalization (or scaling) is different. Since the justification of this is completely analogous to  that of (the first-order result  of)~\cite[Theorem~2]{TahBloch17}, we omit the proof for the sake of brevity.
\end{enumerate}

 \subsection{Two Degraded BCs}\label{sec:degraded}
We now consider two specific classes of  degraded BCs and show that   modifications of standard techniques are applicable in establishing the outer bound to $\calL_\delta$.  
\begin{corollary} \label{cor:bsc} 
Suppose that $W=\mathrm{BSC}(p_1)$ and $V=\mathrm{BSC}(p_2)$ with $0\le p_1\le p_2\le 1/2$ (without loss of generality)~\cite[Example~5.3]{elgamal}. Then $\calL_\delta$ is as in~\eqref{eqn:covert_cap_reg} with 
\begin{equation}
L_j^* =  (\log\rme )(1-2p_j)   \log\frac{1-  p_j }{p_j} \cdot \sqrt{\frac{2}{\chi_2(Q_1\|Q_0)}},\quad j\in\{1,2\}. \label{eqn:Lj_bsc}
\end{equation}
\end{corollary}
An converse proof of this result follows from Mrs.\ Gerber's lemma~\cite{WZ73} and is presented in Section~\ref{sec:prf_bsc}. This result generalizes~\cite[Example~3]{WangWornellZheng16}.

We now consider the scenario in which the BC $P_{Y_1,Y_2,Z|X}$ consists of three additive white Gaussian  noise  (AWGN) channels~\cite[Sec.~5.5]{elgamal}, i.e.,
\begin{align}
Y_j &=X + \Psi_j,\quad j \in \{1,2\} \label{eqn:awgn2}\\
Z &= X+ \Psi_3,\label{eqn:awgn1}
\end{align}
where $\Psi_1,\Psi_2$, and~$\Psi_3$ are independent zero-mean Gaussian noises with variances $N_1, N_2,$ and $\sigma^2$ respectively. There is no (peak, average, or long-term) power constraint on the codewords~\cite[Sec.~V]{WangWornellZheng16}. Let the ``no communication'' input symbol be $0 \in\bbR$, so by~\eqref{eqn:awgn1}, $Q_0$ is distributed as a zero-mean Gaussian with variance $\sigma^2$. 
\begin{corollary} \label{cor:gauss} 
Suppose that $W$ and $V$ be AWGN channels as in~\eqref{eqn:awgn2} with noise variances satisfying $0\le N_1\le N_2$ (without loss of generality). Then $\calL_\delta$ is as in~\eqref{eqn:covert_cap_reg} with 
\begin{equation}
L_j^* =   \frac{\sigma^2\log\rme }{N_j},\quad j\in\{1,2\}. \label{eqn:Lj_gauss}
\end{equation}
\end{corollary}
The converse    proof   follows from the entropy power inequality~\cite{Shannon48,StamEPI59} and is presented in Section~\ref{sec:prf_gauss}. This result generalizes~\cite[Theorem~6]{Bloch16} and~\cite[Theorem~5]{WangWornellZheng16}. It is also analogous to~\cite[Prop.~1]{LTU03} in which it was shown that time-division is optimal for Gaussian BCs in the low-power limit.

We note that Corollaries~\ref{cor:bsc}   and~\ref{cor:gauss}  apply to an arbitrary  but finite number  of {\em successively degraded} legitimate receivers~\cite[Sec.~5.7]{elgamal}, say $N$. This means that $X\markov Y_1\markov Y_2\markov\ldots\markov Y_N$.  The corresponding $\delta$-covert capacity region is 
\begin{equation}
\calL_\delta= \bigg\{ (L_1, \ldots, L_N)\in\bbR_+^N: \sum_{j=1}^N \frac{L_j}{L_j^*} \le 1  \bigg\} , \label{eqn:arb_num}
\end{equation}
where $L_j^*$ is given as in~\eqref{eqn:Lj_bsc} or~\eqref{eqn:Lj_gauss}. 
We omit the  proofs as they are straightforward generalizations of the corollaries.

\subsection{On the Length of the Secret Key} \label{sec:key}
In the preceding derivations, we have assumed that the secret key  that the transmitter and legitimate receivers  share is arbitrarily long. In other words, the set $\calK$ is sufficiently large.  In this section, we derive fundamental limits on the length of the  key  so that covert communication remains successful. 
 
To formalize this, we will need to augment Definition~\ref{def:region}. 
We say that $(L_1, L_2, L_{\mathrm{key}})$ is {\em  $\delta$-achievable} or simply {\em achievable}  if in addition to \eqref{eqn:rate_cons} and \eqref{eqn:rel_constr} (with $\eps=0$),
\begin{equation}
\limsup_{n\rightarrow \infty} \frac{1}{\sqrt{n\delta}} \log K_n\leq L_{\mathrm{key}}.
\end{equation}
Finally, we set $L_Z^*:=L^*(P_{Z,Z|X})$.   The generalization of Theorem~\ref{thm:main} is as follows.
\begin{theorem} \label{thm:key}
Under the conditions of Theorem \ref{thm:main},  the tuple $(L_1, L_2, L_{\mathrm{key}}) \in\bbR_+^3$ is achievable if  and only if 
\begin{align}
\frac{L_1}{L_1^*}+\frac{L_2}{L_2^*}&\leq 1  \label{eqn:ts_opt}
\end{align}
and 
\begin{equation}
 L_{\mathrm{key}} \geq \left(\frac{L_1}{L_1^*}+\frac{L_2}{L_2^*}\right)L_Z^*-L_1-L_2 . \label{eqn:key_ach}
\end{equation}
\end{theorem}
Note that if the throughputs of the code $(\log M_{1n},\log M_{2n})$ are such that we operate on the boundary of the covert capacity region (i.e., that~\eqref{eqn:ts_opt} holds with equality), the optimum (minimum) key length 
\begin{equation}L_{\mathrm{key}}^*=L_Z^*- L_1 -L_2.
\end{equation}
 Also note that if $L_{\mathrm{key}}$ is sufficiently large, then~\eqref{eqn:key_ach} is satisfied so based on~\eqref{eqn:ts_opt},  Theorem~\ref{thm:key} reverts to Theorem~\ref{thm:main}. 
The proof of this enhanced theorem follows largely along the same lines as that for Theorem~\ref{thm:main}. However, we need to carefully bound the length of the secret key. The additional arguments to complete the proof of Theorem~\ref{thm:key} are provided in Section~\ref{sec:prf_key}.

\section{Proof of Theorem~\ref{thm:main}} \label{sec:prf_thmmain} 
\subsection{Preliminaries for the Proof of Theorem~\ref{thm:main}}
Before we commence, we recap some basic notions in convex analysis. The {\em (upper) concave envelope} of $f:\calD\to\bbR$, denoted as  $\frakC[f] (x) := \inf\{g(x): g\ge f \mbox{ over }\calD, g \mbox{ is concave}\}$, is  the smallest concave function lying above $f$. If $\calD$ is a subset of $\bbR^d$, then by Carath\'eodory's theorem~\cite{Nair2013},
\begin{equation}
g=\frakC[f]\quad\Longleftrightarrow\quad g(x) = \sup_{ \{(x_i , p_i)\}_{i=1}^{d+1}} \sum_{i=1}^{d+1} p_i f(x_i), \label{eqn:cave_finite}
\end{equation}
where $x_i\in\calD$ and $\{p_i\}_{i=1}^{d+1}$ is a probability distribution such that $\sum_{i=1}^{d+1}p_ix_i=x$. 
Here, we record a basic fact: 
\begin{equation}
f_2\ge f_1 \;\;\mbox{on}\;\;\calD \quad\Longrightarrow\quad g_2\ge g_1\; \;\mbox{on}\;\;\calD \label{eqn:ce_mono}
\end{equation}
where $g_j = \frakC[f_j], j = 1,2$. 
This can be   shown by means of the representation of the concave envelope in~\eqref{eqn:cave_finite}. Indeed,
\begin{align}
g_2(x)
&\ge  \sum_{i=1}^{d+1} p_i f_2(x_i) \ge \sum_{i=1}^{d+1} p_i f_1(x_i)  \label{eqn:g1g2}
\end{align}
where the first inequality  holds for any $\{(x_i,p_i)\}_{i=1}^{d+1}$ such that $\sum_{i=1}^{d+1}p_ix_i=x$ and the second inequality because $f_2\ge f_1$ on $\calD$. Since the inequality holds for all $\{(x_i,p_i)\}_{i=1}^{d+1}$ such that $\sum_{i=1}^{d+1}p_ix_i=x$, we can take the supremum of the right-hand-side of  \eqref{eqn:g1g2} over all such $\{(x_i,p_i)\}_{i=1}^{d+1}$ to conclude that $g_2\ge g_1$ on $\calD$. 

\subsection{Converse Proof of Theorem~\ref{thm:main}: Binary-Input BCs} \label{subsec:converse} 
 We first prove the converse to Theorem \ref{thm:main} for the case when $\calX=\{0,1\}$. This is done for the sake of clarity and simplicity.  We subsequently show how to extend the analysis to the multiple symbol case (i.e., $|\calX|>2$) in Section \ref{sec:multiple_input}.   
Fix a sequence of $(n,M_{1n},M_{2n},\eps_n,\delta)$-codes for the  BC with a warden $P_{Y_1,Y_2,Z|X}$  satisfying the $0$-reliability constraint in~\eqref{eqn:rel_constrain} and~\eqref{eqn:rel_constr} and the covertness constraint in~\eqref{eqn:convert_constraint}. In the proof, we use the following result by Bloch \cite[Lemma~1, Remark~1]{Bloch16}: 
\begin{lemma} \label{lemma:weight} 
Let $Q_{\gamma}(z):=\sum_{x\in\calX} P_{Z|X}(z|x) P_{\gamma}(x)$ for $\gamma\in [0,1]$ where  $P_\gamma$ is defined in~\eqref{eqn:Palpha_n}.  
Then, it follows that 
\begin{align}
I(P_{\gamma} , P_{Z|X})=\gamma D(Q_1\| Q_0) - D(Q_{\gamma}\|Q_0). \label{eqn:I_gamma}
\end{align}
Furthermore,  for any sequence $\gamma_n$ such that $\gamma_n\to 0$ as $n\to\infty$, for all  $n$ sufficiently large, 
\begin{align}
&\frac{\gamma_n^2}{2}\chi_2(Q_1\|Q_0) \left(1 -\sqrt{\gamma_n} \right)  \le D(Q_{\gamma_n} \| Q_0)\nn\\*
&\qquad\qquad\qquad\qquad\le\frac{\gamma_n^2}{2}\chi_2(Q_1\|Q_0) \left(1 + \sqrt{\gamma_n} \right). \label{eqn:bd_bloch}
\end{align}

\end{lemma}

\subsubsection{Covertness Constraint} \label{subsub:covert_binary}
We first discuss the covertness constraint in~\eqref{eqn:convert_constraint}. Let $Z^n$ (resp. $X^n$) have distribution $\hatQ_{Z^n}$ (resp. $\hatP_{X^n}$) and let $\hatQ_{Z_i}$ (resp. $\hatP_{X_i}$)  be the marginal of $\hatQ_{Z^n}$ (resp. $\hatP_{X^n}$) on the $i$-th element. Additionally, let $\barQ_n$ be the {\em average output distribution} on $\calZ$, i.e., $\barQ_n:=\frac{1}{n}\sum_{i=1}^n \hatQ_{Z_i}$. Similarly we define the {\em average input distribution} $\barP_n$ on $\calX$ as $\barP_n :=\frac{1}{n}\sum_{i=1}^n \hatP_{X_i}$. Then mimicking the steps in the proof of~\cite[Theorem~1]{WangWornellZheng16}, we have 
\begin{align}
D( \hatQ_{Z^n}\| Q_0^{\times n}) 
&\ge nD( \barQ_n \| Q_0). \label{eqn:divergence}
\end{align}
Thus, by the covertness constraint  in~\eqref{eqn:convert_constraint} and~\eqref{eqn:divergence}, we have
\begin{equation}
D( \barQ_n \| Q_0)\le\frac{\delta}{n}. \label{eqn:bound_warden}
\end{equation}
Since $\lim_{n\rightarrow 0} D( \barQ_n \| Q_0)=0$ and symbol $0\in\calX$ is not redundant,\footnote{Indeed, if  $0$ {\em were} redundant (e.g., $P_{Z|X}(\cdot|0) = P_{Z|X}(\cdot|1)$ for binary $\calX$), there exists an input distribution $\tilP$ such that $\tilP(0)=0$ and its corresponding output distribution $\tilQ=\sum_x \tilP(x) P_{Z|X}(\cdot|x)=Q_0$~\cite[Eqn.~(5)]{WangWornellZheng16}. The distribution $\tilQ$ (taking the role of $\barQ_n$) satisfies $D(\tilQ\|Q_0)=0$ so~\eqref{eqn:bound_warden} is trivially satisfied. However, $\tilP$ (taking the role of $\barP_n$) clearly does not satisfy~\eqref{eqn:input_distr}.}   it follows that 
\begin{align}
\barP_n=P_{\alpha_n}  \mbox{ for some }\alpha_n\rightarrow 0. \label{eqn:input_distr}
\end{align}
Because $\barQ_n(z)=Q_{\alpha_n}(z)$ for all $z\in\calZ$, by Lemma \ref{lemma:weight},  
\begin{align}
&\frac{\alpha_n^2}{2}\chi_2(Q_1\|Q_0) \left(1 -\sqrt{\alpha_n} \right)\le D(\barQ_{n} \| Q_0)\nn\\*
&\qquad\qquad\qquad\qquad\le\frac{\alpha_n^2}{2}\chi_2(Q_1\|Q_0) \left(1 + \sqrt{\alpha_n} \right). \label{eqn:bd_bloch_specialized}
\end{align}
From~\eqref{eqn:bound_warden} and~\eqref{eqn:bd_bloch_specialized},  we conclude that $\alpha_n$ has  to satisfy the weight constraint
\begin{equation}
\alpha_n^2 \left(1-\sqrt{\alpha_n} \right)\le {\frac{2\delta}{\chi_2(Q_1\|Q_0) n}}=:\bar{\alpha}_n^2.\label{eqn:wt_const}
\end{equation} From this relation, we see that $\alpha_n \le \bar{\alpha}_n (1+o(1))=\Theta(\frac{1}{\sqrt{n}})$. 

\subsubsection{Upper Bound  on Linear Combination of Code Sizes} \label{subsub:linear_binary}
We now proceed to consider upper bounds on the code sizes subject to the reliability and covertness constraints.  Without loss of generality, we assume $L_1^*\ge L_2^*$. We start with a lemma that is a direct consequence of the converse proof for more capable BCs by El Gamal~\cite{elgamal79}. This lemma is stated in a slightly different manner in~\cite[Theorem~8.5]{elgamal}. 
\begin{lemma} \label{lem:conv}
Every $(n,M_{1n},M_{2n}, \eps_n)$-code  for any BC satisfies 
\begin{align}
(\log M_{1n})(1-\eps_n)-1&\le  \sum_{i=1}^n I(U_{1i} ;Y_{1i}),\label{eqn:conv_bdM1} \\
(\log M_{2n})(1-\eps_n) -1 &\le  \sum_{i=1}^n I(U_{2i} ;Y_{2i}),\label{eqn:conv_bdM2} 
\end{align}\vspace{-.3in}
\begin{align}
&(\log M_{1n}+\log M_{2n} )(1-\eps_n)-2  \nn\\*
&\qquad\le  \sum_{i=1}^n  \big[ I(X_i;Y_{1i} |U_{2i} )+  I(U_{2i} ;Y_{2i})\big] ,\label{eqn:conv_bdsum1} \\
&(\log M_{1n}+\log M_{2n} )(1-\eps_n)-2  \nn\\*
&\qquad\le  \sum_{i=1}^n  \big[ I(U_{1i} ;Y_{1i})+I(X_i;Y_{2i} |U_{1i}) \big], \label{eqn:conv_bdsum2}
\end{align}
where $U_{1i} :=( W_1,Y_1^{i-1}, Y_{2,i+1}^n)$ and $U_{2i} := (W_2, Y_{1,i+1}^n,Y_2^{i-1})$ satisfies $(U_{1i}, U_{2i}) \markov  X_i \markov  (Y_{1i}, Y_{2i})$.  In addition, if a secret key $S$ is available to the encoders and decoder, then   the auxiliary random variables $U_{1i}$ and $U_{2i}$ also include $S$.
\end{lemma}
For completeness, the proof of Lemma~\ref{lem:conv}, with the effect of the secret key, is provided in Appendix~\ref{app:prf_conv}. Note that no assumption (e.g., degradedness, less noisy or more capable conditions) is made on the BC in Lemma~\ref{lem:conv}. 

Now fix a constant $\lambda\ge 1$.  By adding $(\lambda-1)$ copies   of~\eqref{eqn:conv_bdM2} to one copy of~\eqref{eqn:conv_bdsum1} and writing $U_i:=U_{2i}$ for all $i\in \{1,\ldots, n\}$, we obtain 
\begin{align}
&(\log M_{1n}  + \lambda \log M_{2n} )(1-\eps_n)-(1+\lambda) \nn\\*
&\qquad\le \sum_{i=1}^n\big[ I(X_i;Y_{1i}|U_i) + \lambda   I(U_i; Y_{2i}) \big]. \label{eqn:bl_lemma0}
\end{align}
By introducing the usual time-sharing random variable $J\in \mathrm{Unif}\{1,\ldots, n\}$ which is independent of all the other random variables $(W_1, W_2, X^n, Y_1^n,Y_2^n)$ and defining $U :=(J, U_J)$, $X:=X_J$, $Y_1:=Y_{1J}$ and $Y_2 := Y_{2J}$, we obtain
\begin{align}
&(\log M_{1n}  + \lambda \log M_{2n} )(1-\eps_n)-(1+\lambda)\nn\\*
&\qquad\le n I(X;Y_1|U) +n\lambda I(U;Y_2) .\label{eqn:bl_lemma}
\end{align}
Note that the above identification of $(U,X)$ satisfies $U\markov X\markov (Y_1,Y_2)$. Furthermore, since $X$ is equal to $X_i$ for each  $ i \in \{1,\ldots, n\}$ with equal probability $1/n$ (because of $X=X_J$), it follows $P_X=\barP_n=P_{\alpha_n}$  for some $\alpha_n$ satisfying~\eqref{eqn:wt_const}. Hence, \eqref{eqn:bl_lemma} can be bounded by taking  a maximization over all such distributions, i.e., 
\begin{align}
&\frac{1 }{n}  \big[(\log M_{1n}  + \lambda \log M_{2n} )(1-\eps_n  )-(1+\lambda)\big] \nn\\*
&\qquad\le  \max_{P_X} \max_{ P_{U|X} } I(X;Y_1|U)+\lambda I (U;Y_2), \label{eqn:max_P}
\end{align}
where the maximization over $P_X$  is  over distributions $P_{\alpha_n}$ where $\alpha_n$ satisfies~\eqref{eqn:wt_const}.

Now by using the concave envelope representation for the capacity region of degraded BCs~\cite{Nair2013}, we may write~\eqref{eqn:bl_lemma} as follows: 
\begin{align}
&\frac{1 }{n}  \big[(\log M_{1n}  + \lambda \log M_{2n} )(1-\eps_n  )-(1+\lambda)\big] \nn\\*
&\;\le  \max_{P_X} \max_{ P_{U|X} } I(X;Y_1|U)+\lambda I (U;Y_2)\label{eqn:linear}\\
 &\;= \max_{P_X}\max_{ P_{U|X} } I(X;Y_1|U)+\lambda[ I (X;Y_2)-I(X;Y_2|U)] \label{eqn:follows_markov}\\ 
 &\;=  \max_{P_X}\lambda  I (X;Y_2) + \max_{P_{U|X} }  I(X;Y_1|U)-\lambda I(X;Y_2|U)\\
  &\;=\max_{P_X}\lambda  I (X;Y_2) + \frakC_{P_X}[ I(X;Y_1 )-\lambda I(X;Y_2 )]  \label{eqn:cave_env}.
\end{align}
Here,  \eqref{eqn:follows_markov} follows from the Markov chain $U\markov X\markov Y_2$ and~\eqref{eqn:cave_env} follows from the definition of  the concave envelope. Namely,  for fixed $P_X$, 
\begin{align}
&\!\!\!\max_{P_{U|X} }  I(X;Y_1|U)-\lambda I(X;Y_2|U) \nn\\*
&\! \!\!=\! \max_{P_{U|X}} \sum_u \! P_U(u) [ I(X;Y_1|U\! =\! u)\!-\!\lambda I(X;Y_2|U\! =\! u)]\\
&\!\!\!=\! \max_{P_{U|X}} \sum_u \! P_U(u) [ I(P_{ X|U}(\cdot|u) , W )\!-\!\lambda I(P_{ X|U}(\cdot|u), V)] \!\label{eqn:use_markovity}\\
&\!\!\!=\!\frakC_{P_X}[ I(X;Y_1 )-\lambda I(X;Y_2 )]  , \label{eqn:cave_opt}
\end{align}
where~\eqref{eqn:use_markovity} follows from the Markov chain $U\markov X\markov (Y_1,Y_2)$ and~\eqref{eqn:cave_opt} follows from the fact that $\sum_u P_{U}(u) P_{X|U}(\cdot|u)=P_X$ and~\eqref{eqn:cave_finite}.  Note that we employ the subscript $P_X$ on $\frakC$ to emphasize that the concave envelope operation is taken with respect to the distribution $P_X$ and it is thus a function of $P_X$.

\subsubsection{Approximating the Maximization over Low-Weight Inputs}
Due to the above considerations, it now suffices to simplify~\eqref{eqn:cave_env}. To obtain the outer bound to~\eqref{eqn:covert_cap_reg}, we set $\lambda:= L_1^*/L_2^*$ in the following. Note that since  it is assumed that $L_1^*\ge L_2^*$, $\lambda\ge 1$ as required to obtain~\eqref{eqn:bl_lemma0}. If instead $L_2^*> L_1^*$, then we set $\lambda:=L_2^*/L_1^*$ and add $(\lambda-1)$ copies of~\eqref{eqn:conv_bdM1} to one copy of~\eqref{eqn:conv_bdsum2} to obtain the analogue of~\eqref{eqn:bl_lemma0} with $U_i:=U_{1i}$. We also use \eqref{eqn:cond2} instead of \eqref{eqn:cond1} in Condition~\ref{cond1}. Finally, in the rest of the proof, we  replace the index~$1$ by~$2$ and~$W$ by~$V$  and vice versa. The following arguments  go through verbatim with these minor amendments. 

By expressing the mutual information quantities in~\eqref{eqn:cave_env} as $I(P_X, W)$ and $I(P_X,V)$~\cite{Csi97} (for $I(X;Y_1)$ and $I(X;Y_2)$ respectively), we can write~\eqref{eqn:cave_env} as follows:
\begin{equation}
\max_{P_X} \lambda  I (P_X,V ) + \frakC_{P_X}[ I (P_X ,W)-\lambda I (P_X ,V )]   \label{eqn:max_alpha} .
\end{equation}

By applying \eqref{eqn:cond1} of Condition~\ref{cond1}, we see that $ I (P_X ,W)-\lambda I (P_X ,V )\le 0$ for all $P_X$. Hence, by using the monotonicity property in~\eqref{eqn:ce_mono}  and the fact that the concave envelope of the $0$ function is $0$, we obtain that~\eqref{eqn:max_alpha} is upper bounded by 
 \begin{equation}
\max_{P_X} \lambda  I (P_X,V )  \label{eqn:alpha_n_D} 
\end{equation} 
Now we  parametrize $P_X = P_{\alpha_n}=\barP_n$ as the vector $[1-\alpha_n,\alpha_n]^T$ where $0\le\alpha_n\le\bar{\alpha}_n(1+o(1))$  [cf.~\eqref{eqn:wt_const}]. Appealing to Lemma~\ref{lemma:weight}, we see that  the value of the   optimization problem in~\eqref{eqn:alpha_n_D}  is given by 
\begin{equation}
\lambda   \bar{\alpha}_n(1+o(1))  \cdot D(V(\cdot | 1 ) \| V(\cdot | 0 )).\label{eqn:alpha_n_D2} 
 \end{equation} 
Now, combining~\eqref{eqn:max_alpha}--\eqref{eqn:alpha_n_D2}  with the upper bound in~\eqref{eqn:cave_env}, we have 
\begin{align}
&\frac{1 }{n} \big[(\log M_{1n}  + \lambda \log M_{2n} )(1-\eps_n)-(1+\lambda)\big]\nn\\*
&\qquad    \le\lambda \bar{\alpha}_n(1+o(1))\cdot  D(V(\cdot | 1 ) \| V(\cdot | 0 )). \label{eqn:combine_both}
\end{align}
\subsubsection{Completing the Converse Proof Taking Limits} \label{subsub:completing}
At this point, we invoke the definition of $\bar{\alpha}_n$ in~\eqref{eqn:wt_const}  and normalize~\eqref{eqn:combine_both}  by $\sqrt{n\delta}$ to obtain 
\begin{align}
&\!\!\!  \frac{1 }{\sqrt{n\delta}}\big[ (\log M_{1n}  + \lambda \log M_{2n} )(1-\eps_n)-(1+\lambda)\big]\nn\\*
&\! \! \!  \le  \lambda (1+ o(1))\sqrt{ \frac{2\delta n}{\chi_2(Q_1\|Q_0)\delta n} } \cdot    D(V(\cdot | 1 ) \| V(\cdot | 0 ))   . \! 
\end{align}
Taking the $\limsup$ in $n$ on both sides, recalling that  (i) $\lambda=L_1^*/L_2^*$ and (ii) the definition of  achievable~$(L_1,L_2)$  pairs according to Definition~\ref{def:region}, and the facts that $\eps_n\to 0 $ and $n\delta\to \infty$, we obtain
\begin{equation}
  L_1+\frac{L_1^*}{L_2^*} L_2\le \frac{L_1^*}{L_2^*} \sqrt{ \frac{2 }{\chi_2(Q_1\|Q_0)}}\cdot  D(V(\cdot | 1 ) \| V(\cdot | 0 ))   .\label{eqn:L1_D0}
\end{equation}
Finally, by recalling the definition of the covert capacity of $P_{Y_2,Z|X}$ according to~\eqref{eqn:Lbin} in  Theorem~\ref{def:cov_cap},  we see that the right-hand-side of~\eqref{eqn:L1_D0} is exactly $L_1^*$. Hence we obtain the desired outer bound corresponding to $\calL_{\delta}$ in~\eqref{eqn:covert_cap_reg} in Theorem~\ref{thm:main}. 



\subsection{Converse Proof of Theorem~\ref{thm:main}:   BCs with Multiple Inputs}\label{sec:multiple_input}
We generalize the converse proof for binary-input in Section~\ref{subsec:converse} to  the multiple symbol  case, i.e., $|\calX|>2$. We use the following lemma whose proof is in  \cite[Section VII-B]{Bloch16}.   
\begin{lemma} \label{lemma:weight_multi} 
For $\gamma\in [0,1]$ and length-$K$ probability vector $\mathbf{p}=[p_1 ,\ldots,  p_K]^T$, i.e., $p_k\geq 0$ and $\sum_{k=1}^K p_k=1$, let $Q_{\gamma, \mathbf{p}}(z):=\sum_{x\in\calX} P_{Z|X}(z|x) P_{\gamma, \mathbf{p}}(x)$ where 
\begin{equation}
P_{\gamma, \mathbf{p}} (x) := \left\{ \begin{array}{cc}
1-\gamma& x = 0 \\
 \gamma p_k & x = k \in  \calX\setminus\{0\} \\
\end{array} \right.. 
\end{equation} 
Then, it follows that
\begin{equation}
\! \!  I(P_{\gamma, \mathbf{p}},P_{Z|X}) \! =\!  \bigg(\sum_{k=1}^K\gamma  p_k D(Q_k\| Q_0) \bigg) - D(Q_{\gamma,\mathbf{p}}\|Q_0). \label{eqn:mi_gen}
\end{equation}
Furthermore,  for any sequence $\gamma_n$ such that $\gamma_n\to 0$ as $n\to\infty$, for all  $n$ sufficiently large, 
\begin{align}
&\frac{\gamma_n^2}{2}\chi_2 \bigg(\sum_{k=1}^K p_k Q_k \bigg\|Q_0 \bigg) \left(1 -\sqrt{\gamma_n} \right)\le D(Q_{\gamma_n, \mathbf{p}} \| Q_0)\nn\\*
&\qquad\qquad\qquad\le\frac{\gamma_n^2}{2}\chi_2 \bigg(\sum_{k=1}^K   p_kQ_k \bigg\|Q_0 \bigg) \left(1 + \sqrt{\gamma_n} \right) \label{eqn:bd_bloch2}
\end{align}
\end{lemma}

\subsubsection{Covertness Constraint} By a close inspection of the steps in  Section~\ref{subsub:covert_binary} and by applying Lemma~\ref{lemma:weight_multi}, we   conclude that the average input distribution $\barP_n$ is given by $P_{\alpha_n,  \mathbf{p}}$ for some length-$K$ probability vector $\mathbf{p}$ and $\alpha_n$ such that 
\begin{equation}
\alpha_n^2 \left(1-\sqrt{\alpha_n} \right)\le {\frac{2\delta}{\chi_2\big( \sum_{k=1}^K  p_k Q_k\big\|Q_0 \big) n}}=:\bar{\alpha}_{\bp, n}^2.\label{eqn:wt_const_gen}
\end{equation}  

\subsubsection{Upper Bound on Linear Combination of Code Sizes} Let us assume $L_1^*\ge L_2^*$ (without loss of generality), then $\lambda = L_1^*/L_2^*\geq 1$.  It can be easily checked that the steps in Section~\ref{subsub:linear_binary} go through in the same manner for the  multiple symbol  case 
except that $P_X=P_{\alpha_n, \bf{p}}$ for some length-$K$ probability vector $\mathbf{p}$ and $\alpha_n\in [0,1]$ satisfying \eqref{eqn:wt_const_gen}. 
 As such, we have that~\eqref{eqn:cave_env} also holds where now the maximization over $P_X$ is over those distributions $P_{\alpha_n,\bp}$ for some $\bp$ and $\alpha_n$ satisfying~\eqref{eqn:wt_const_gen}.  

\subsubsection{Approximating the Maximization over Low-Weight Inputs} Now we approximate \eqref{eqn:max_alpha}. We fix $\bp$ throughout.  By Condition~\ref{cond1}, $I(P_X,W)-\lambda I(P_X,V)\le 0$ and so only the first term $\lambda I(P_X,V)$ remains in the maximization. Similarly to~\eqref{eqn:alpha_n_D2}, the maximization in \eqref{eqn:max_alpha}  can be approximated by 
\begin{equation}
\lambda \bar{\alpha}_{\bp,n} (1+o(1))  \sum_{k=1}^K p_k D(V(\cdot | k ) \| V(\cdot | 0 ))
\end{equation}
as $n\to\infty$.

Due to the same reasoning up to \eqref{eqn:combine_both}, we have 
\begin{align}
&\!\!\!\frac{1 }{n} \big[(\log M_{1n}  + \lambda \log M_{2n} )(1-\eps_n)-(1+\lambda)\big]\nn\\*
&\!\!\!\le \lambda \bar{\alpha}_{\bp,n}(1 + o(1))\cdot 	\bigg(\sum_{k=1}^K p_k D(V(\cdot |k ) \| V(\cdot | 0 )) \bigg)   \label{eqn:combine_both_multi}.
\end{align}

\subsubsection{Completing the Converse Proof Taking Limits} By substituting $\lambda=L_1^*/L_2^*$ and $\bar{\alpha}_{\bp,n}$ defined in~\eqref{eqn:wt_const_gen} into \eqref{eqn:combine_both_multi},  taking limits as in Section \ref{subsub:completing}, and subsequently maximizing over all probability vectors $\bp$ on $\calX\setminus \{0\}$, we obtain 
\begin{align}
L_1+\frac{L_1^*}{L_2^*} L_2 &\le\frac{L_1^*}{L_2^*} \max_{\mathbf{p} }\sqrt{\frac{2 \big(\sum_{k=1}^Kp_k D(V(\cdot |k ) \| V(\cdot | 0 )) \big)^2}{\chi_2 \big(\sum_{k=1}^K p_k Q_k\big\|Q_0\big)}} \nn\\*
&= L_1^*.
\end{align}
This completes the proof for the multiple symbol  case.

\section{Proofs of Corollaries in Section~\ref{sec:degraded}} \label{sec:prf_coro}
\subsection{Proof of Corollary~\ref{cor:bsc} } \label{sec:prf_bsc}
 As the the capacity region of the stochastically degraded BC is the same as that of the physically degraded BC with the same marginals \cite[Sec. 5.4]{elgamal}  even under the covertness constraint, we assume that the physically degradedness condition holds, i.e., $Y_1=X\oplus \Psi_1, Y_2=Y_1\oplus \tilde{\Psi}_2$ where $\tilde{\Psi}_2\sim \mathrm{Bern}(\tilde{p}_2)$ such that $p_1\ast\tilp_2=p_2$. 

 We follow the exposition in~\cite[Sec.~5.4.2]{elgamal} closely.  By the standard argument to establish the weak converse for degraded BCs~\cite{Gal74}, we know that every $(n,M_{1n}, M_{2n},\eps_n$)-code for a BC must satisfy
\begin{align}
( \log M_{1n} )(1-\eps_n)-1 &\le n I(X;Y_{1 } | U) \label{eqn:bound_M1} \\
( \log M_{2n} )(1-\eps_n)-1 &\le n I(U ;Y_{2 }  ) ,\label{eqn:bound_M2}
\end{align}
for some $U\markov X\markov (Y_1,Y_2)$ in which 
  $X$ satisfies the weight constraint in~\eqref{eqn:wt_const}, i.e., $P_X=P_{\alpha_n}$ or  $\bbE[X]=\alpha_n$. 
We now evaluate the region in~\eqref{eqn:bound_M1}--\eqref{eqn:bound_M2}    explicitly in terms of the channel parameters and the upper bound on the weight of $X$. 

Now we have  
\begin{align}
h_{\rmb}(p_2) &=  H(Y_2 |X) \\
&=  H(Y_2 |U,X)\le H(Y_2|U) \le H(Y_2) \label{eqn:UX}\\
&=  h_{\rmb}(\alpha_n \ast p_2) ,\label{eqn:hp2}
\end{align}
where the equality in~\eqref{eqn:UX} follows from the Markov chain $U \markov X\markov Y_2$ and~\eqref{eqn:hp2} follows from the fact that the entropy of $Y_2 = X\oplus \Psi_1\oplus \tilde{\Psi}_2$ when  $\bbE [X] = \alpha_n$ is $h_{\rmb}( \alpha_n\ast p_2)$. As a result, from~\eqref{eqn:hp2}, there exists a  $\tau_n\in [0,\alpha_n]$ such that 
\begin{equation}
H(Y_2 |U)= h_{\rmb}(\tau_n \ast p_2).
\end{equation}
Then,  we have 
\begin{align}
h_{\rmb}(\tau_n \ast p_2) &=H(Y_2|U) = H(Y_1 \oplus \tilde{\Psi}_2 |U)\\
& \ge h_{\rmb} \big(h_{\rmb}^{-1} ( H(Y_1 |U)  ) \ast \tilp_2 \big) \label{eqn:mrs_gerbers}
\end{align}
where~\eqref{eqn:mrs_gerbers} follows from the conditional version of  Mrs.\ Gerber's lemma~\cite{WZ73}. By the monotonicity of $h_{\rmb}(\cdot)$ on $[0,1/2]$ and the fact that $p_1\ast\tilp_2 = p_2$, we have 
\begin{equation}
\tau_n \ast p_1\ast\tilp_2  \ge  h_{\rmb}^{-1} ( H(Y_1 |U)  ) \ast \tilp_2 .
\end{equation}
As a result, we have 
\begin{equation}
H(Y_1 |U) \le h_{\rmb}( \tau_n \ast  p_1).
\end{equation}
Now consider
\begin{align}
I(X;Y_1|U) &= H(Y_1|U) - H(Y_1 | U,X)\\*
& = H(Y_1|U)-h_{\rmb}(p_1)\\*
&\le  h_{\rmb}( \tau_n \ast p_1)-h_{\rmb}(p_1).\label{eqn:bound_M1s}
\end{align}
For the other rate bound in~\eqref{eqn:bound_M2}, we have 
\begin{align}
I(U;Y_2)& = H(Y_2) - H(Y_2|U) \\*
&\le h_{\rmb}( \alpha_n \ast p_2) - h_{\rmb}(\tau_n\ast p_2).\label{eqn:bound_M2s}
\end{align}
Now, note that for all $q \in (0,1/2)$, $\frac{\rmd }{\rmd q}h_{\rmb} (q) =  (\log\rme)\log\frac{1-q}{q}$. By Taylor expanding $h_{\rmb}$ around $q$, we obtain
\begin{align}
&h_{\rmb}(q\ast \xi)   = h_{\rmb}( q + (1-2q) \xi)  \\*
&\qquad=  h_{\rmb}(q)+ (\log\rme)\log\frac{1-q}{q}\cdot (1-2q) \xi +O(\xi^2) \label{eqn:taylor}
\end{align}
as $\xi\to 0$. 
Now we can simplify the bounds on the right-hand-sides of~\eqref{eqn:bound_M1s} and~\eqref{eqn:bound_M2s}.  
Write $\alpha_n := a/\sqrt{n}$ and $\tau_n:=t/\sqrt{n}$ where $t\in [0,a]$ and $a:=(1+o(1))\sqrt{ 2\delta/\chi_2(Q_1\|Q_0)}$; see~\eqref{eqn:wt_const} for the relation between $\alpha_n$ and an upper bound on a function of it given by $\bar{\alpha}_n$. By combining~\eqref{eqn:bound_M1},~\eqref{eqn:bound_M2},~\eqref{eqn:bound_M1s},~\eqref{eqn:bound_M2s}, and~\eqref{eqn:taylor},  we obtain the outer bound
\begin{align}
&\frac{1}{n} \big[( \log M_{1n} )(1-\eps_n)-1 \big] \nn\\*
&\quad\le (\log\rme)(1-2p_1) \log\frac{1-p_1}{p_1} \cdot \frac{t}{\sqrt{n}}  + O\left(\frac{1}{n} \right)  \label{eqn:bound_M1ss} \\* 
&\frac{1}{n} \big[( \log M_{2n} )(1-\eps_n)-1 \big]\nn\\*
&\quad\le  (\log\rme)(1-2p_2) \log\frac{1-p_2}{p_2} \cdot \frac{a-t}{\sqrt{n}}+ O\left(\frac{1}{n} \right)  .\label{eqn:bound_M2ss}
\end{align}
By dividing the first and second bounds by $(\log\rme)(1-2p_1) \log\frac{1-p_1}{p_1}$ and $(\log\rme)(1-2p_2) \log\frac{1-p_2}{p_2}$ respectively, adding  them, multiplying the resultant expression by $\sqrt{n/\delta}$, and taking limits, we immediately recover the outer bound to $\calL_\delta$ with $L_j^*$ defined in~\eqref{eqn:Lj_bsc}. 
\subsection{Proof of Corollary~\ref{cor:gauss} } \label{sec:prf_gauss}
The proof follows similarly to that of Corollary~\ref{cor:bsc} but we use the entropy power inequality~\cite{Shannon48,StamEPI59} in place of Mrs.\ Gerber's lemma. As in the proof of Corollary~\ref{cor:bsc},  we assume   physically degradedness, i.e., $Y_1=X+\Psi_1, Y_2=Y_1+\tilde{\Psi}_2$ where $\tilde{\Psi}_2$ is independent zero-mean Gaussian noise with variance $N_2-N_1$. 
Let the second moment of the input  distribution be denoted as $\alpha_n=\bbE[X^2]$; this plays the role of the weight of $X$ in the discrete case. The bounds~\eqref{eqn:bound_M1} and~\eqref{eqn:bound_M2} clearly still hold so we only have to single-letterize the two mutual information terms $I(X;Y_{1 } | U) $ and $I(U;Y_2)$. We follow the exposition in~\cite[Sec.~5.5.2]{elgamal} closely. 

First, we have 
\begin{align}
I(U;Y_2)&=h(Y_2)-h(Y_2|U)\\*
&\le \frac{1}{2}\log(2\pi \rme(\alpha_n+N_2)) -h(Y_2|U) \label{eqn:gauss_max}
\end{align}
where~\eqref{eqn:gauss_max} follows from the fact that a Gaussian maximizes differential entropy over all distributions with the same second moment.
Now note that 
\begin{align}
\frac{1}{2}\log(2\pi \rme N_2)&=  h(Y_2| X) \\
&=h(Y_2| U,X)\le h(Y_2|U)\le h(Y_2) \\
&\le \frac{1}{2}\log(2\pi \rme(\alpha_n+N_2)).
\end{align}
As such there exists a $\tau_n\in [0,\alpha_n]$ such that
\begin{equation}
h(Y_2|U)=\frac{1}{2}\log (2\pi\rme (\tau_n+N_2)).
\end{equation}
Then, by the conditional form of the entropy power inequality~\cite{Shannon48,StamEPI59},
\begin{align}
h(Y_2|U) &=h(Y_1+\tilde{\Psi}_2|U)\\
&\ge \frac{1}{2}\log \left(2^{2h(Y_1|U)}+2^{2 h(\tilde{\Psi}_2|U)} \right)\\
&= \frac{1}{2}\log \left(2^{2h(Y_1|U)}+2\pi\rme(N_2-N_1)\right).
\end{align}
Thus,
\begin{equation}
h(Y_1|U)\le \frac{1}{2}\log(2\pi\rme(\tau_n+N_1)).
\end{equation}
Now, we are ready to upper bound the mutual information terms using the above calculations. We have
\begin{align}
I(X;Y_1|U)&=h(Y_1|U)-h(Y_1|U,X)\\
&\le  \frac{1}{2}\log(2\pi\rme(\tau_n+N_1))-\frac{1}{2}\log (2\pi\rme N_1) \\
&=\frac{1}{2}\log \left( 1+\frac{\tau_n}{N_1}\right) \le \frac{\tau_n\log\rme }{2N_1} ,\label{eqn:first_mi_gauss}
\end{align}
where the last inequality follows from the fact that $\log(1+t)\le t \log\rme $ for $t\ge -1$. 
Similarly, 
\begin{align}
I(U;Y_2)&=h(Y_2)-h(Y_2|U ) \\
&\le \frac{1}{2}\log \left(\frac{\alpha_n+N_2}{\tau_n+N_2}\right)\le \frac{(\alpha_n-\tau_n)\log\rme }{2(\tau_n+N_2)}\\
&\le\frac{(\alpha_n-\tau_n)\log\rme }{2N_2}.\label{eqn:sec_mi_gauss}
\end{align}
Using the same calculations as those leading to~\cite[Eqn.~(75)]{WangWornellZheng16}, we conclude that the covert communication constraint in the Gaussian case translates to 
\begin{equation}
\frac{\alpha_n^2}{4\sigma^4}+o(\alpha_n^2)\le\frac{\delta}{n}. \label{eqn:wt_gauss}
\end{equation}
The proof is completed by uniting~\eqref{eqn:first_mi_gauss},~\eqref{eqn:sec_mi_gauss}, and~\eqref{eqn:wt_gauss} in a way that is analogous    to the conclusion of the proof  for the binary symmetric BC in Section~\ref{sec:prf_bsc}. 

\section{Proof Sketch of Theorem~\ref{thm:key}} \label{sec:prf_key}
\begin{IEEEproof}
The achievability in~\eqref{eqn:key_ach} follows by   augmenting  the standard time-division argument given in  Remark~1 in Section~\ref{sec:remarks}. 
More specifically, we let  $\rho_1:={L_1}/{L_1^*}$ and $\rho_2:= {L_2}/{L_2^*}$. Then $\rho_1+\rho_2\leq 1$. We use an optimal code for the first receiver for $\rho_1$ fraction of time, an optimal code for the second receiver for $\rho_2$ fraction of time, and stay idle for the remaining fraction of time. Then the required secret key rate\footnote{With a slight abuse of terminology, we will refer to $L_1, L_2$, and $L_{\mathrm{key}}$ as {\em rates} even though they are not  communication rates in the usual sense~\cite{elgamal}.}   is $\rho_1(L_Z^*-L_1^*)+\rho_2(L_Z^*-L_2^*)$ as a direct consequence of~\cite[Theorem~3]{Bloch16}, which is equal to the right-hand side of~\eqref{eqn:key_ach}.

For the converse, for simplicity, we only consider the binary-input case; the general case follows from replacing $Q_1$   with $\sum_{k=1}^K p_kQ_k$ everywhere and $\barP_n(1)$ with $\sum_{k=1}^K\barP_n(k)$ everywhere. We show that there is a tradeoff between the message  rates and the secret key rate  (with normalizations $\sqrt{n\delta}$); if the message rates are in the interior of~\eqref{eqn:covert_cap_reg}, a strictly smaller secret key rate  is required. To elucidate this tradeoff, let us consider a rate tuple $(L_1, L_2)$ such that $L_1/L_1^*+L_2/L_2^*=\kappa$ for some $\kappa\in [0,1]$. Note that $\kappa$ strictly less than one means that the message rates are in the interior of~\eqref{eqn:covert_cap_reg}. 

By applying similar steps as in \cite[Eqns.~(91)--(94)]{Bloch16}, we obtain
\begin{align}
 &\log M_{1n} +\log M_{2n}+\log K_n= H(W_1, W_2, S)\\
&\geq I(W_1, W_2, S; Z^n) 
\geq I(X^n; Z^n) 
 \geq nI(X;Z)-\delta,\label{eqn:use_hou}
\end{align}
where the final inequality in~\eqref{eqn:use_hou} follows from similar steps presented in~\cite[Sec.~5.2.3]{hou_thesis}. Here $X=X_J$ and  $Z=Z_J$ where $J$ is the time-sharing random variable defined in Section~\ref{subsub:linear_binary}. Then, 
\begin{align}
& \frac{1}{\sqrt{n\delta}}(\log M_{1n}+\log M_{2n}+\log K_n)\nn\\*
&\quad\geq \frac{1}{\sqrt{n\delta}}(nI(X;Z)-\delta) \\
& \quad= \frac{n\alpha_n}{\sqrt{n\delta}}\left(D(Q_1\|Q_0)-\frac{D(Q_{\alpha_n}\|Q_0)}{\alpha_n}-\frac{\delta}{n\alpha_n}\right),  \label{eqn:use_bloch_lemma}
\end{align}
where~\eqref{eqn:use_bloch_lemma} follows from Lemma \ref{lemma:weight}. Note that the lower bound \eqref{eqn:use_bloch_lemma} depends on the 
``weight'' $\alpha_n =\barP_n(1)$. Hence, substituting the smallest ``weight'' $\alpha_n =\barP_n(1)$ that attains  $L_1/L_1^*+L_2/L_2^*=\kappa$ into \eqref{eqn:use_bloch_lemma}  results in  a lower bound on the required secret key rate to attain the message rate  pair~$(L_1, L_2)$.  

Now let us denote the smallest  $\alpha_n =\barP_n(1)$ that attains  $L_1/L_1^*+L_2/L_2^*=\kappa$ by  $\nu \bar{\alpha}_n$.\footnote{We omit mulitplying $\bar{\alpha}_n$ by the factor $1+o(1)$ as this is inconsequential in the (asymptotic) arguments that follow.}  By a close inspection of the proof of Theorem~\ref{thm:main}, we can check that under the constraint $\alpha_n= \nu  \bar{\alpha}_n$, the message pairs are bounded by  $L_1/L_1^*+L_2/L_2^*\leq \nu $. Hence, to achieve  $L_1/L_1^*+L_2/L_2^*=\kappa$, it should follow that $\nu\geq \kappa$. Indeed, we can see that $L_1/L_1^* +L_2/L_2^*=\kappa$ is achievable with weight $\kappa \bar{\alpha}_n$ from the direct part above,  by communicating $\kappa$ fraction of the time and staying idle for the remaining $1-\kappa$ fraction. 

Note that when $\alpha_n=\kappa \bar{\alpha}_n$, it  follows from \eqref{eqn:bd_bloch} that 
\begin{align}
\lim_{n\rightarrow \infty} \frac{D(Q_{\alpha_n}\|Q_0)}{\alpha_n}=0 \quad\mbox{and}\quad 
\lim_{n\rightarrow \infty} \frac{\delta}{n\alpha_n} =0. \label{eqn:lim2}
\end{align}
Then, by combining \eqref{eqn:use_bloch_lemma}--\eqref{eqn:lim2}, we have 
\begin{align}
&\liminf_{n\rightarrow \infty }\frac{1}{\sqrt{n\delta}}(\log M_{1n}+\log M_{2n}+\log K_n)\nn\\*
&\qquad\geq \kappa \sqrt{\frac{2D(Q_1\|Q_0)^2}{\chi_2(Q_1\|Q_0) }} =\kappa L_Z^*. \label{eqn:s_key}
\end{align}
Since $\kappa= {L_1}/{L_1^*}+{L_2}/{L_2^*}$,~\eqref{eqn:s_key} immediately implies a  matching  converse to~\eqref{eqn:key_ach}.\end{IEEEproof}

\section{Conclusion and Future Work} \label{sec:concl}
In this paper, we established the covert capacity region for two-user memoryless BCs that satisfy Condition~\ref{cond1}. Somewhat surprisingly, the most  basic multi-user communication strategy---time-division transmission---turns out to be optimal for this class of BCs. Our proof strategy provides further evidence  that the concave envelope characterization of bounds on capacity regions in network information theory~\cite{Nair2013} is   convenient and  useful.

There are  at least three promising avenues for future work. 
\begin{enumerate}[leftmargin=*]
\item Because we are adopting the {\em average} probability of error formalism in~\eqref{eqn:rel_constrain}, it is likely that the strong converse (in $\varepsilon$) does not hold as suggested by the argument in Appendix~A of~\cite{TahBloch17}. Verifying that the argument therein indeed extends to BCs would be a natural avenue for future work. In addition, proving that a strong converse holds under the {\em maximum}  probability of error formalism would also be a fruitful endeavor.  

\item What about the covert capacity region for BCs that do not satisfy Condition~\ref{cond1}? In this case time-division may not be optimal and a natural avenue for future work would be to construct schemes to beat the time-division inner bound for such BCs. 
\item What is the covert capacity region for BCs (under appropriate conditions) in which there are more than two legitimate receivers? While Corollaries~\ref{cor:bsc} and~\ref{cor:gauss} hold for an arbitrary number of legitimate receivers (see~\eqref{eqn:arb_num}), Theorem~\ref{thm:main}, and in particular Lemma \ref{lem:conv} in which it hinges on,  does not seem to generalize easily.  
\end{enumerate}
\appendices
\section{Proof of Lemma~\ref{lem:conv}} \label{app:prf_conv}
\begin{IEEEproof}
We will only prove~\eqref{eqn:conv_bdM1} and~\eqref{eqn:conv_bdsum1} as the other two bounds follow by swapping indices~$1$ with~$2$ and vice versa. In addition, in this proof, we include the effect of the secret key $S$.  By Fano's inequality,
\begin{align}
H(W_j|Y_j^n,S)& \le \eps_n\log M_{jn} +1,\quad j \in \{1,2\}.
\end{align}
Because $W_1$ and $W_2$ are uniform and independent of each other and of $S$,  we have 
\begin{align}
( \log M_{1n} )(1-\eps_n)-1&\le  I(W_1;Y_1^n|S) \label{eqn:bl_1} \\
( \log M_{1n} +\log M_{2n} )(1-\eps_n)-2 &\le I(W_1;Y_1^n|W_2,S) \nn\\*
&\quad+ I(W_2;Y_2^n|S).\label{eqn:bl_2}
\end{align}
We start by single letterizing~\eqref{eqn:bl_1} as follows:
\begin{align}
&( \log M_{1n} )(1-\eps_n)-1\nn\\* 
&\le  \sum_{i=1}^n I(W_1; Y_{1i} | Y_1^{i-1},S)  \\
&\le  \sum_{i=1}^n I(W_1,S,Y_1^{i-1},Y_{2,i+1}^n; Y_{1i} ) \\
&= \sum_{i=1}^n I(U_{1i}; Y_{1i} ) ,\label{eqn:U1}
\end{align}
where~\eqref{eqn:U1} follows from the identification $U_{1i} := (W_1, S,Y_1^{i-1},Y_{2,i+1}^n)$.  This proves~\eqref{eqn:conv_bdM1}. 

Next we single letterize~\eqref{eqn:bl_2} using the steps at the top of the next page,
\begin{figure*}[h]
\begin{align}
&( \log M_{1n} +\log M_{2n} )(1-\eps_n)-2  \nn\\*
&\le\sum_{i=1}^n I(W_1; Y_{1i} | W_2, Y_{1,i+1}^n,S) + I(W_2;Y_{2i}| Y_2^{i-1},S) \\
 &\le \sum_{i=1}^n I(W_1, Y_2^{i-1}; Y_{1i} | W_2,S, Y_{1,i+1}^n) + I(W_2,  S,Y_2^{i-1};Y_{2i} ) \\
  &= \sum_{i=1}^n I(W_1, Y_2^{i-1}; Y_{1i} | W_2,S,Y_{1,i+1}^n ) + I(W_2,  S,Y_{1,i+1}^n, Y_2^{i-1};Y_{2i}  )  - I(Y_{1,i+1}^n;Y_{2i} | W_2,S, Y_2^{i-1})\\
    &= \sum_{i=1}^n I(W_1; Y_{1i} | W_2, S,Y_{1,i+1}^n,  Y_2^{i-1}) + I(Y_2^{i-1}; Y_{1i} | W_2,S,Y_{1,i+1}^n)  \nn\\*
    &\qquad+ I(W_2, S,Y_{1,i+1}^n, Y_2^{i-1};Y_{2i}  ) - I(Y_{1,i+1}^n;Y_{2i} | W_2,  S,Y_2^{i-1})\\
    &= \sum_{i=1}^n I(W_1; Y_{1i} | U_{2i} ) + I( U_{2i}; Y_{2i}) \label{eqn:csi}\\
        &\le \sum_{i=1}^n I(X_i; Y_{1i} | U_{2i} ) + I( U_{2i}; Y_{2i}),\label{eqn:x_func}
\end{align}\makebox[\linewidth]{\rule{2\columnwidth}{0.4pt}}
\end{figure*} 
where~\eqref{eqn:csi} follows from the Csisz\'ar-sum-identity~\cite{elgamal} and the identification $U_{2i} := (W_2,S,Y_{1,i+1}^n, Y_2^{i-1})$ and~\eqref{eqn:x_func} follows from the fact that $X_i$ is a function of $(W_1,W_2,S)$. This proves~\eqref{eqn:conv_bdsum1}. 
\end{IEEEproof}
%

\subsection*{Acknowledgements}   
The authors thank Dr.\ Ligong Wang (ETIS---Universit\'e Paris Seine, Universit\'e de Cergy-Pontoise, ENSEA, CNRS) for helpful discussions in the initial phase of this work and for pointing us to~\cite{LTU03}.  The first author also thanks Dr.\ Lei Yu (NUS) for fruitful discussions on the relation of the present work to stealth~\cite{hou_thesis}.

The authors would like to thank the Associate Editor, Prof.\ Rainer B\"ohme, and the anonymous reviewers for their useful feedback to improve the quality of the manuscript.


\begin{IEEEbiography}[{\includegraphics[width=1in,height=1.25in,clip,keepaspectratio]{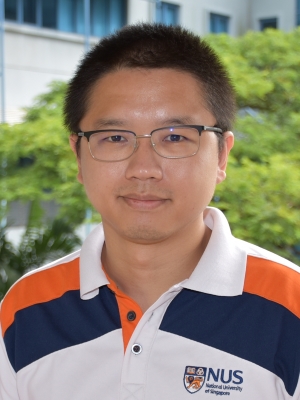}}] {Vincent Y.\ F.\ Tan} (S'07-M'11-SM'15)  was born in Singapore in 1981. He is currently an Associate Professor in the Department of Electrical and Computer Engineering  and the Department of Mathematics at the National University of Singapore (NUS). He received the B.A.\ and M.Eng.\ degrees in Electrical and Information Sciences from Cambridge University in 2005 and the Ph.D.\ degree in Electrical Engineering and Computer Science (EECS) from the Massachusetts Institute of Technology (MIT)  in 2011.  His research interests include information theory, machine learning, and statistical signal processing.

Dr.\ Tan received the MIT EECS Jin-Au Kong outstanding doctoral thesis prize in 2011, the NUS Young Investigator Award in 2014, the NUS Engineering Young Researcher Award in 2018, and the Singapore National Research Foundation (NRF) Fellowship (Class of 2018). He is also an IEEE Information Theory Society Distinguished Lecturer for 2018/9. He has authored a research monograph on {\em ``Asymptotic Estimates in Information Theory with Non-Vanishing Error Probabilities''} in the Foundations and Trends in Communications and Information Theory Series (NOW Publishers). He is currently serving as an Associate Editor of the IEEE Transactions on Signal Processing.
\end{IEEEbiography}

\begin{IEEEbiography}[{\includegraphics[width=1in,height=1.25in,clip,keepaspectratio]{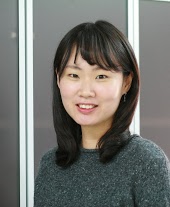}}] {Si-Hyeon Lee}  (S'08-M'13)
 is an Assistant Professor in the Department of Electrical Engineering at the Pohang University of Science and Technology (POSTECH), Pohang, South Korea. She received the B.S. (summa cum laude) and Ph.D. degrees in Electrical Engineering from the Korea Advanced Institute of Science and Technology (KAIST), Daejeon, South Korea, in 2007 and 2013, respectively. From 2014 to 2016, she was a Postdoctoral Fellow in the Department of Electrical and Computer Engineering at the University of Toronto, Toronto, Canada. Her research interests include network information theory, physical layer security, and wireless communication systems.

\end{IEEEbiography}
\end{document}